\documentclass[onecolumn,12pt]{autartarxiv}
\usepackage{graphicx}
\usepackage{comment}
\usepackage{arydshln}
\usepackage{mathtools}
\usepackage{graphics} 
\usepackage{epsfig} 
\usepackage{times} 
\usepackage{amsmath} 
\usepackage{amssymb}  
\usepackage[mathscr]{eucal}
\usepackage{enumerate}
\usepackage{mathrsfs}
\usepackage{multirow}
\usepackage{subfigure}
\usepackage{latexsym}
\usepackage{bm}
\usepackage{dsfont}
\usepackage{multicol}
\usepackage{multirow}
\usepackage{color}
\usepackage{url}
\usepackage{algorithm}
\usepackage{cite}
\usepackage{algpseudocode}
\newtheorem{remark}{Remark}
\newtheorem{assumption}{Assumption}
\newtheorem{lemma}{Lemma}
\newtheorem{proposition}{Proposition}
\newtheorem{definition}{Definition}

\newtheorem{theorem}{Theorem}

\newcommand{\xl}[1]{{\color{black}  { #1}}}

{
	\begin{bmatrix}}%
	{\end{bmatrix}
	}

\def\build#1_#2^#3{\mathrel{\mathop{\kern0pt#1}\limits_{#2}^{#3}}}%

\def\min#1{\build{\rm min}_{#1}^{}}
\def\build#1_#2^#3{\mathrel{\mathop{\kern0pt#1}\limits_{#2}^{#3}}}%

\begin{document}
	
	\begin{frontmatter}
		
		\title{Robust Tube-based Model Predictive Control  with Koopman Operators--\emph{Extended Version}} 
		
		\author[XLZ]{Xinglong Zhang}\ead{zhangxinglong18@nudt.edu.cn},    
		\author[WP]{Wei Pan}\ead{wei.pan@tudelft.nl},               
		\author[RS]{Riccardo Scattolini}\ead{riccardo.scattolini@polimi.it},  
		\author[SY]{Shuyou Yu}\ead{shuyou@jlu.edu.cn},               
		\author[XLZ]{Xin~Xu}\ead{xinxu@nudt.edu.cn}  
		\address[XLZ]{College of Intelligence Science and Technology, National University of Defense Technology, Changsha 410073, China.}  
		\address[WP]{Department of Cognitive Robotics, Delft University of Technology, the Netherlands.	}             
		\address[RS]{Dipartimento di Elettronica, Informazione e Bioingegneria, Politecnico di Milano, Milan 20133, Italy.}
		\address[SY]{Department of Control Science and Engineering, Jilin University
			at NanLing, Changchun 130025, China.}       


		\begin{abstract}                          
Koopman operators are of infinite dimension and capture the characteristics of nonlinear dynamics in a lifted global linear manner. The finite data-driven approximation of Koopman operators results in a class of linear predictors,  useful for formulating linear model predictive control (MPC) of nonlinear dynamical systems with reduced computational complexity. However, the robustness of the closed-loop Koopman MPC under modeling approximation errors and possible exogenous disturbances is still a crucial issue to be resolved.
Aiming at the above problem, this paper presents a  robust tube-based MPC solution with Koopman operators, i.e., r-KMPC, for nonlinear discrete-time dynamical systems with additive disturbances.  The proposed controller is composed of a nominal MPC using a lifted Koopman model and an off-line nonlinear feedback policy.  The proposed approach does not assume the convergence of the approximated Koopman operator, which allows using a Koopman model with a limited order for controller design. Fundamental properties, e.g., stabilizability, observability,  of the Koopman model are derived under standard assumptions with which, the closed-loop robustness and nominal point-wise convergence are proven. 
Simulated examples are illustrated to verify the effectiveness of the proposed approach.
		\end{abstract}
		
	\end{frontmatter}
	
\section{Introduction}
Model predictive control (MPC),  is employed as an effective control tool for control of numerous applications, such as robotics and industrial plants, see~\cite{mayne2000constrained,qin2003survey}. 
%
In MPC, a prediction model is typically required, with which many MPC algorithms can be formulated, e.g., stabilizing MPC for nominal models in~\cite{rawlings2009model} and the references therein, and robust MPC such as min-max MPC in~\cite{bemporad2003min} or tube-based MPC in~\cite{mayne2005robust,mayne2011tube} for systems under model uncertainties and disturbances. Focusing on building well-performed prediction models, designing MPC using data-driven models has been noted as a promising direction.
Many works on this aspect  have been developed, see~\cite{piga2017direct,carron2019data,terzi2018learning,peitz2020data,bujarbaruah2019adaptive,kohler2019linear} and the references therein. 
Among them,  a unitary learning-based predictive controller for linear systems was addressed in~\cite{terzi2018learning}, where  set membership is adopted to estimate a multi-step linear prediction model to be used for designing a robust MPC. Similarly, resorting to the set membership identification, adaptive MPC algorithms for uncertain time-varying systems were proposed in~\cite{lorenzen2017adaptive,fagiano2015scenario}  for reducing the conservativity caused by robust MPC. Relying on the main idea from iterative learning control, a data-driven learning MPC for repetitive tasks was studied in~\cite{rosolia2017learning} with terminal constraints updated iteratively. In these approaches, a linear robust (or stabilizing) MPC problem is to be solved online since the considered model is linear. \\
In the control of nonlinear systems, the derivation of a nonlinear prediction model could be a nontrivial task. In~\cite{limon2017learning}, a nonlinear MPC algorithm using a machine learning-based model estimation was proposed with stability property guarantees. However, nonlinear MPC results in nonlinear, even nonconvex, optimization problems, which can be computationally intensive for systems with high nonlinearity. To reduce the computational load, a supervised machine learning algorithm was used to approximate the nonlinear MPC law in~\cite{hertneck2018learning}. The robustness was guaranteed under bounded control approximation errors with verified statistic empirical risk. \\
%
%
In another line, Koopman operators have been noted to be effective to represent the internal dynamics of nonlinear systems~\cite{arbabi2018data,korda2018convergence}. 
Specifically, the Koopman operator is typically with infinite dimension, capturing the nonlinear dynamical characteristics through a linear dynamic evolution on a lifted observable function of states. In~\cite{korda2018linear}, a finite-dimensional truncation of the Koopman operator was used to form a linear predictor of nonlinear dynamics for designing a linear MPC. This approach paves the way to linear MPC formulations of nonlinear systems with a linear predictor that represents a wide operation range. Among the most related works, the extension to an MPC algorithm using the integrated Koopman operator was addressed in~\cite{peitz2020data}, and MPC methods using a deep learning-based Koopman model were developed in~\cite{lian2021koopman,han2020deep}. The application of Koopman MPC to flow control was addressed in~\cite{arbabi2018data}. \xl{In~\cite{narasingam2020data}, a  Lyapunov-based Koopman MPC was developed for the feedback stabilization of nonlinear systems.} \\
Note that, the presence of modeling errors is almost inevitable in  lifted Koopman models, due to the data-driven finite-dimensional approximation  of Koopman operators and  exogenous disturbances, see~\cite{korda2018linear,williams2015data}. \xl{An interesting offset-free Koopman MPC extension of~\cite{narasingam2020data}  was presented in~\cite{son2020handling} for handling model-mismatch with an estimator.  However, the satisfaction of hard state constraints and closed-loop robustness under both non-negligible approximating errors and unknown additive disturbances are crucial concerns, which were not addressed in the previous Koopman-based MPC~\cite{son2020handling,narasingam2020data,korda2018linear,peitz2020data,lian2021koopman}.} This motivated our research work.\\
In this paper, we present a robust MPC solution with Koopman operators in the framework of tube-based MPC~\cite{mayne2005robust,mayne2011tube}.
{\color{black}The contributions are twofold.
The first contribution is a linear robust Koopman-based  MPC design methodology  for nonlinear systems with unknown dynamics and additive disturbances. As opposed to the classic Tube-based MPC~\cite{mayne2005robust,mayne2011tube},  our approach allows designing robust MPC from measured data and no explicit model information is required; also, our approach results in a nonlinear MPC law by a linear robust MPC design. 
The second contribution is the analysis of the closed-loop theoretical  guarantees for Koopman-based MPC under modeling errors and additive disturbances.  This is achieved via imposing standard prior conditions on the lifting observable functions, allowing for using a truncated Koopman model with a limited system order in the controller design.}\\
%
The rest of the paper is organized as follows.  Section 2 introduces the considered control problem and preliminary solutions. In Section 2 the main idea of the proposed r-KMPC and the associated theoretical results are obtained. Section 4 shows the simulation results obtained by applying the proposed approach to nontrivial simulated systems, while some conclusions are drawn in Section 5. The ingredients for estimating the uncertainty terms are given in the Appendix.\\
\textbf{Notation:} We denote $\mathbb{N}$ as the set of positive natural numbers and $\mathbb{N}_{1}^l$ as the numbers $1,\cdots,l$. 
Given the variable $r$, we use  ${\bm r}_{\scriptscriptstyle k:k+N}$ to denote the sequence $r_k\ldots r _{k+N}$ and $\bm r_k$ to denote ${\bm r}_{\scriptscriptstyle k:k+N}$ after its first appearance, where $k$ is the discrete time index and $N$ is a positive integer. For a vector $x\in\mathbb{R}^{n}$, we use $\|x\|_Q^2$ to stand for $x^{\top}Qx$, $\|x\|$ to denote its Euclidean norm; while for a matrix $x\in\mathbb{R}^{n\times m}$, we denote $\|x\|_F$ as the Frobenius norm.  Given two sets $\mathcal{Z}$ and $\mathcal{V}$, their Minkowski sum  is represented by $\mathcal{Z}\oplus \mathcal{V}=\{z+v|z\in \mathcal{Z}, v\in \mathcal{V}\}$. 
For a given set of variables $z_{i}\in{\mathbb{R}}^{q_{i}}$,
$i=1,2,\dots,M$, we define the vector whose vector-components are $z_{i}$ in the following compact form: $(z_{1}, z_{2}, \cdots, z_{\rm\scriptscriptstyle M})=[\,z_{1}^{\top}\ z_{2}^{\top}\ \cdots\ z_{\rm\scriptscriptstyle M}^{\top}\,]^{\top}\in{\mathbb{R}}^{q}$, where $q= \sum_{i=1}^{M}q_{i}$.
\section{Control problem and preliminaries}
\subsection{Control problem}
Consider a class of nonlinear discrete-time  systems with additive disturbances described by
\begin{equation}\label{Eqn:non-model}
x^+=f(x,u)+w_o,
\end{equation}
where $x\in \mathcal{X}\subset \mathbb{R}^{n}$, $u\in \mathcal{U}\subset \mathbb{R}^m$ are the state and control variables, $x^+$ is the successor state at the next discrete-time instant, $w_o\in\mathcal{W}_o$ is an additive bounded noise which can be unknown and not measurable, $\mathcal{W}_o$ is a compact set containing the origin,  $\mathcal{X}$ and $\mathcal{U}$ are convex sets containing the origin in their interiors, $f$ is the state transition function, which can be partially or completely unknown. It is assumed that $f(0,0)=0$,  $f(x,u)$ is $C^{\infty}$ on $\mathcal{X}\times \mathcal{U}$, and $\|f(x,u)\|<+\infty$ for all ${x\times u\in\mathcal{X}\times \mathcal{U}}$.  
The state $x$ is measurable.\\ 
Starting from any initial condition $x_0\in\mathcal{X}$, the control objective is to minimize a quadratic cost of type $J=\sum_{k=0}^{+\infty} \|x_k\|_{Q}^2+\|u_k\|_R^2,$
where  $Q=Q^{\top}\in\mathbb{R}^{n\times n}$ and $R=R^{\top}\in\mathbb{R}^{m\times m}$, $Q,R\succ 0$.\\

\begin{definition}[Local stabilizability~\cite{bacciotti1988local}]\label{Eqn:cost-finite}
	Model $x_{k+1}=f(x_k,u_k)$ is stabilizable on the domain $\mathcal{X}\times \mathcal{U}$ if, for any $x_0\in\mathcal{X}$, there exists a 
	feedback function $u(x_k)\in\mathcal{U}$, $u(0)=0$, such that state $x_k$ of the corresponding closed-loop system asymptotically converges to the origin. 
\end{definition}

\begin{definition}[Generalized gradient~\cite{clarke1975generalized}]
	The generalized gradient of a Lipschitzian function $h(x):\mathbb{R}^{n}\rightarrow\mathbb{R}^{q}$ at $x$, denoted as $\triangledown h(x)$, is the convex hull of all matrices of the form $M=\lim_{i\rightarrow \infty}\triangledown h(x+\delta x_i)$, where $\delta x_i\rightarrow 0$ as $i\rightarrow\infty$.
\end{definition}

\begin{definition}[Maximal rank~\cite{clarke1976inverse}]
	The generalized gradient $\triangledown h(x)$ is of maximal rank, if for every $M$   ${\rm{rank}}(M)=\break\min\,  \{n,q\}$.
\end{definition}
\subsection{Preliminary Koopman MPC}

{\color{black}
We first review the Koopman operator theory for autonomous dynamical systems and its extension to dynamical systems with controls.\\ 
%
%
%
 Let us introduce the so-called observable of $x$ defined by a scalar-valued function $\phi(x):\mathcal{X}\rightarrow \mathbb{C}$ and  denote $\mathcal{F}$ as a given space of observables. 
 For an autonomous model $x^+=f(x,0)$, i.e. model~\eqref{Eqn:non-model} with $u,w_o=0$,
the Koopman operator $\mathcal{K}:\mathcal{F}\rightarrow \mathcal{F}$ is defined by~\cite{arbabi2018data,korda2018convergence} 
\begin{equation}\label{Eqn:koopman_auto}
\mathcal{K} \phi(x)=\phi(x)\circ f(x,0), 
\end{equation}
for every observable $\phi(x)\in\mathcal{F}$ ($\mathcal{F}$ is invariant under the action of the Koopman operator), $\circ$ is the composition operator, i.e., $\mathcal{K} \phi(x)=\phi(f(x,0))$. For any  discrete-time instant $k\in\mathbb{N}$, it holds that
 \begin{equation*}
\phi(x_k)=\mathcal{K}\phi(x_{k-1})=\cdots=\mathcal{K}^k\phi(x_0),
 \end{equation*}
%
 which captures the dynamical characteristics of the original nonlinear dynamics. For a detailed introduction of the Koopman operator please refer to~\cite{klus2020kernel,klus2020data,korda2018convergence}.\\
%
The Koopman operator for $x^+=f(x,0)$ can be generalized to systems with controls (i.e. model~\eqref{Eqn:non-model}) in several ways,  see e.g.~\cite{williams2016extending,proctor2018generalizing,korda2018linear}. In our study, we adopt the practical and rigorous scheme in~\cite{korda2018linear}, which relies upon an extended state space $\mathcal{X}\times \ell(\mathcal{U\times W}_o)$, where $\ell(\mathcal{U\times W}_o)$ is the space of all the sequences composed of the control and disturbance, i.e. $\boldsymbol{u}_w:=\{{u}_{w}(i)\}_{i=0}^{+\infty}$ with $u_w(i)=(u(i),w_o(i))\in\mathcal{U}\times \mathcal{W}_o$. Letting $f_W(x,u_w):=f(x,u)+w_o$ and $\boldsymbol{u}_w(i)$ be the $i$-th element of $\boldsymbol{u}_w$, 
  one can write the dynamics of the extended state   $\boldsymbol \chi=(x,\boldsymbol{u}_w)$ 
as 
\begin{equation}\label{Eqn:extended-model}
\boldsymbol \chi^+=F(\boldsymbol \chi):=(f_W(x,\boldsymbol{u}_w(0)), {\varGamma} \boldsymbol{u}_w),
\end{equation}
where $\varGamma$ is a left shift operator such that $\boldsymbol{u}_w(i+1)=\varGamma\boldsymbol{u}_w(i)$. In this way,  the Koopman operator $\mathcal{K}:\mathcal{F}_e\rightarrow \mathcal{F}_e$ associated with~\eqref{Eqn:extended-model} is given by~\cite{korda2018linear}
\begin{equation}\label{Eqn:koopman_ex}
\mathcal{K} \phi(\boldsymbol \chi)=\phi(\boldsymbol \chi)\circ F(\boldsymbol \chi)
\end{equation}
where $\phi({\boldsymbol \chi}):\mathcal{X}\times\ell(\mathcal{U\times W}_o)\rightarrow \mathbb{C}$ belongs to the extended observable space $\mathcal{F}_e$, which contains observable functions on arguments $x$ and $\boldsymbol{u}_w$. \\
A finite-dimensional numerical approximation of $\mathcal{K}$ in~\eqref{Eqn:koopman_ex} is of interest for controller design, which can be computed by resorting to the extended dynamic mode decomposition (EDMD) method in a data-driven manner, see~\cite{korda2018linear}.
Let a finite-dimensional approximation of $\mathcal{K}$ be $\mathcal{K}_{\scriptscriptstyle N_{\phi}}\in\mathbb{R}^{N_{\phi}\times N_{\phi}}$, associated with an observable vector $\Phi(\boldsymbol\chi)=(\phi_1(\boldsymbol \chi),\cdots,\phi_{N_{\phi}}(\boldsymbol \chi))$. The goal in EDMD is to compute $\mathcal{K}_{\scriptscriptstyle N_{\phi}}$ via minimizing $
\|\Phi(\boldsymbol \chi^+)-\mathcal{K}_{\scriptscriptstyle N_{\phi}}\Phi(\boldsymbol \chi)\|^2.$
 Note however that $\boldsymbol \chi=(x,\boldsymbol{u}_w)$ is of infinite-dimension, which can be problematic from the computational viewpoint. 
Hence, we choose a computable observable function as
\begin{equation}\label{Eqn:lift-extend}
\Phi(x,\boldsymbol u_w)=\left(
\Psi(x),
\boldsymbol{u}_w(0)\right),%
\end{equation}
where $N_{\phi}=n_{\psi}+m+n$, $n_{\psi}>n$, 
 $$\Psi\left(x\right) :=( {\psi_{1}\left(x\right)},{\cdots}, {\psi_{n_{\psi}}\left(x\right)}),$$
and $\psi_i$, $i\in\mathbb{N}_{1}^{n_{\psi}}$, can be chosen as some basis functions or neural networks~\cite{lian2021koopman}. 
%
 \xl{ To compute $\mathcal{K}_{\scriptscriptstyle N_{\phi}}$ with EDMD, let us assume to have collected  $M$ input-state datasets $\{(u_i,\hat w_{o,i}, x_i,x_i^+)\}_{i=1}^M$ satisfying $x_i^{+}=f(x_i,u_i)+w_{o,i}$, where  $\hat w_{o,i}$ is the estimation of $w_{o,i}$ which can be computed by resorting to a nonlinear estimation technique or  a 
 	Koopman operator-based estimator~\cite{surana2016linear}}. \\
 \xl{The following condition is assumed to hold~\cite{korda2018convergence,klus2020kernel,klus2020data}.
 	\begin{assumption}\label{Eqn:data-distribution}
  The data points $\{(u_i,x_i)\}_{i=1}^M$ are drawn independently according to a non-negative probability distribution $\mu$. 

 \end{assumption}
This condition can also be replaced with the assumption that $\{(u_i,x_i)\}_{i=1}^M$ are ergodic in $\mathcal{U}\times\mathcal{X}$ with respect to $\mu$, which can be  generated by integrating a stochastic dynamics, see~\cite{korda2018convergence,klus2020kernel}.} \\ 
  Let $[\mathcal{K}_{\scriptscriptstyle N_{\phi}}]_{\scriptscriptstyle 1:n_{\psi}}=[A\, B\, D]\in\mathbb{R}^{n_{\psi}\times n_{\phi}}$, where $A\in \mathbb{R}^{n_{\psi}\times n_{\psi}}$, $B\in \mathbb{R}^{n_{\psi}\times m}$, and $D\in \mathbb{R}^{n_{\psi}\times n}$. Since we are only interested in predicting $\Psi(x^+)$ with $\Psi(x)$ and the estimation of $\boldsymbol{u}_w(0)$ (due to $\hat w_o$), 
  the  following regularized least squares problem can be stated: \\ 
\begin{equation}\label{Eqn:appro_K}
\begin{array}{ll}
\min{[\mathcal{K}_{\scriptscriptstyle N_{\phi}}]_{\scriptscriptstyle 1:n_{\psi}}} \sum_{i=1}^{M}\|[\mathcal{K}_{\scriptscriptstyle N_{\phi}}]_{\scriptscriptstyle 1:n_{\psi}}(
\Psi(x_i),
\hat {u}_{w,i})-\Psi(x_i^+)\|^2+\alpha\|[\mathcal{K}_{\scriptscriptstyle N_{\phi}}]_{\scriptscriptstyle 1:n_{\psi}}\|_F^2
:=V_{\mathcal{K}}
\end{array}
\end{equation}
where $\hat u_{w,i}=(u_i,\hat w_{o,i})$, $\alpha>0$ is a tuning parameter on the Frobenius norm regularization of $[\mathcal{K}_{\scriptscriptstyle N_{\phi}}]_{\scriptscriptstyle 1:n_{\psi}}$.} 
To recover $x$ using $\Psi(x)$, a linear matrix $C\in\mathbb{R}^{n\times n_{\psi}}$ is  optimized according to the following  problem~\cite{korda2018linear,folkestad2020extended,williams2015data}:
\begin{equation}\label{Eqn:appro_C}
\min{C} \sum_{i=1}^{M}\|C{\Psi(x_i)}-x_i\|^2+\beta\|C\|_F^2,
\end{equation}
where $\beta>0$ is a tuning parameter. 
{\color{black}
By solving~\eqref{Eqn:appro_K} and~\eqref{Eqn:appro_C}, a linear Koopman predictor of~\eqref{Eqn:non-model} can be obtained, i.e.,
\begin{equation}\label{Eqn:unpert}
\left\{\begin{array}{l}
\hat{s}^+=A\hat{s}+B\hat u\\
[0.2cm] \hat x=C\hat s.
\end{array}\right. \qquad 
\end{equation}
Note that, in~\eqref{Eqn:unpert}, a new  (abstract) variable $\hat s$ serves as the state due to the lifted observable construction, while the predicted value of the original state, i.e., $\hat x$, becomes the output variable through the mapping matrix $C$ from the observable space. \\
With~\eqref{Eqn:unpert}, a linear Koopman MPC (KMPC) problem similar to~\cite{korda2018linear} can be stated as follows: 
\begin{equation}\label{Eqn:optimiz0}
\min{{\bm { u}}_{\scriptscriptstyle k:k+N-1}}  \sum_{i=0}^{N-1}(\|\hat {x}_{k+i}\|_{Q}^2+\| u_{k+i}\|_{R}^2)
+\|\hat {x}_{k+N}\|_{Q_N}^2\vspace{-0mm}
\end{equation}
subject to model~\eqref{Eqn:unpert} with $\hat {s}_k=\Psi(x_k)$, state constraint $C\hat {s}_{k+i}\in   {\mathcal{X}},\,\,\forall i\in\mathbb{N}_{1}^{\scriptscriptstyle N}$, and control constraint $u_{k+i}=\hat u_{k+i}\in { {\mathcal{U}}},\,\,\forall i\in\mathbb{N}_{0}^{\scriptscriptstyle N-1}$, where $Q_N=Q_N^{\top}\in\mathbb{R}^{n\times n}$ and $Q_N\succ 0$.\\
	Because of the characteristics of Koopman operators, a merit of KMPC lies in the linear property of the built model~\eqref{Eqn:unpert}, leading to a linear MPC problem instead of a nonlinear one. 
	{\color{black}However, the derivation of~\eqref{Eqn:unpert} with~\eqref{Eqn:lift-extend} by~\eqref{Eqn:appro_K} and~\eqref{Eqn:appro_C} could bring modeling errors,  see~\cite{korda2018linear,folkestad2020extended,williams2015data}, whose property and possible (negative) influences on the closed-loop control performance is not yet analyzed.} As a consequence, the closed-loop property under modeling errors and additive disturbances remains still a crucial issue, which was not addressed in
	the prescribed Koopman-based MPC~\cite{son2020handling,narasingam2020data,korda2018linear,peitz2020data,lian2021koopman}. Peculiarly, with KMPC in~\cite{korda2018linear}, the constraint satisfaction $x\in{\mathcal{X}}$ might not be fulfilled by $C\hat s\in\mathcal{X}$ and the closed-loop robustness of the Koopman MPC might not be verified under modeling errors and disturbances.  
Aiming at this problem, in the following section we propose a robust tube-based MPC using Koopman operators with theoretical guarantees.  

\section{Robust Koopman MPC}
%
In this section, the proposed robust MPC solution using Koopman operators, i.e., r-KMPC, is presented.  First, a Koopman model with approximation errors is derived and its stabilizability and observability properties are proven under standard assumptions. Then, the proposed r-KMPC algorithm using the Koopman model is presented. Finally, the closed-loop theoretical properties of r-KMPC are proven.\\
{\subsection{Koopman model for robust MPC}\label{sec:model-skmpc}
%
%
\xl{As described in~\cite{korda2018linear}, 
 it is not guaranteed that $\mathcal{K}_{\scriptscriptstyle N_{\phi}}$ converges to  $\mathcal{K}$ as $N_{\phi},M\rightarrow +\infty$ even if $w_{o}$ is measurable and $\{w_{o,i}\}_{i=1}^M$ are ergodic samples, because the adopted observable function~\eqref{Eqn:lift-extend} does not form an orthonormal basis of $\mathcal{F}_e$.
Hence, the presence of modeling errors of~\eqref{Eqn:unpert} with~\eqref{Eqn:lift-extend} is inevitable also due to the existence of estimation errors of $w_o$ and  to the practical design with $\alpha,\beta\neq 0$ in~\eqref{Eqn:appro_K} and~\eqref{Eqn:appro_C}. 
Nonetheless, as pointed out in~\cite{otto2021koopman}, the model structure like~\eqref{Eqn:unpert} is of interest from the control viewpoint because it could still be effective for approximating~\eqref{Eqn:non-model} in a large state space region; also it permits a linear MPC implementation for nonlinear systems, leading to a computational load reduction compared with the approaches using complex bilinear models and switched models~\cite{peitz2019koopman,peitz2020data}.  The consideration of using  bilinear or switched models is left as further investigation. \\
To derive a robust Koopman MPC solution with~\eqref{Eqn:unpert},  the fundamental boundedness property of the overall uncertainty caused by multiple sources is required.} To this end, letting ${s}=\Psi(x)$, one can first write an equivalent Koopman model of~\eqref{Eqn:non-model} considering the effects of model uncertainties, that is 
\begin{equation}\label{Eqn:linear_p-residual}
\left\{\hspace{-1mm}\begin{array}{l}
{s}^+=A{s}+Bu+\bar w(s,u,w_o,\hat w_o)\\
[0.2cm] x=C{s}+v(s),
\end{array}\right. \qquad
\end{equation}
where 
$\bar w=D\hat w_o+w(s,u,w_o,\hat w_o)\in\bar {\mathcal{W}}$, $w(s,u,w_o,\hat w_o)\in\mathcal{W}$ and $v(s)\in \mathcal{V}$ are the modeling errors, where $\mathcal{W}$ and $\mathcal{V}$ are convex sets containing the origin; $\bar {\mathcal{W}}=D \hat{\mathcal{W}}_o\oplus \mathcal{W}$, $\hat{\mathcal{W}}_o$ is a computable convex set where $\hat w_{o}$ lies in. \xl{It is assumed that $\hat{\mathcal{W}}_o$ is bounded.
A discussion on the boundedness property of sets $\bar{\mathcal{W}}$ (i.e., $\mathcal{W}$) and $\mathcal{V}$  is deferred to Proposition~\ref{prop:uncertainty-bound}. 
To  this end, we first introduce the following assumption about $\Psi(x)$.} 

{\color{black}
	\begin{assumption}\label{assum:invers-phi}\hfill
		\begin{enumerate}[(1)]
			\item  	The lifted function $\Psi(x)$ is Lipschitz continuous. 
			\item\label{assum3:independent}  $\{\psi_i(x)\}_{i=1}^{n_{\psi}}$ are linearly independent.
		\end{enumerate}
	\end{assumption}
	The verification of the above assumption is easy since many adopted  basis functions (BF) such as  Gaussian kernel functions, polyharmonic splines, and thinplate splines are in fact $C^{\infty}$ and linearly independent.
	\begin{lemma}[Existence of inverse maps~\cite{clarke1976inverse}]\label{lemma:inver-map}
	Letting $\mathcal{S}_{\Psi}$ be the set such that $\mathcal{S}_{\Psi}=\{s\in\mathbb{R}^{n_{\psi}}|s=\Psi(x), x\in\mathcal{X}\}$, if $\triangledown \Psi(x)$ is of maximal rank, i.e., ${\rm{rank}}(\triangledown \Psi(x))=n$, there  exists a Lipschitzian function $\Psi^{-1}:\mathcal{S}_{\Psi}\rightarrow\mathcal{X}$ such that $\Psi^{-1}(\Psi(x))=x$ for every $x\in\mathcal{X}$.
\end{lemma}
%
%
In the following proposition, taking a special type of Gaussian kernels as an example of basis functions, we show indeed that  multiple solutions of $\Psi^{-1}$ can be found. 
\begin{proposition}[Multiple choices of $\Psi^{-1}$]\label{Eqn:multiple-choice}
	Letting $\psi_i(x)=e^{-\|x-c_i\|^2}$, $c_i\in\mathcal{X}$, $i=\mathbb{N}_{1}^{n_{\psi}}$,  a group of any $n+1$ basis functions of $\Psi(x)$ can surely define a choice of $\Psi^{-1}$ if and only if  Assumption~\ref{assum:invers-phi}.(2) holds. 
\end{proposition}
\textbf{Proof}. Since $\psi_i(x)$, $i\in\mathbb{N}_{1}^{n_{\psi}}$ are Gaussian kernels, the resulting $\Psi$ is continuous differentiable. In this case the generalized gradient $\triangledown \Psi(x)$ coincides with the exact gradient $\partial \Psi(x)$.
For a generic $n\in\mathbb{N}$, select a group of any $n+1$ basis functions such that $\|x-c_i\|^2=-\text{log}\psi_i(x)$, $\forall i\in\mathbb{N}_{1}^{n+1}$. Letting $\bar{\psi}(x)=(\psi_1(x),\cdots,\psi_{n+1}(x))$,  the corresponding gradient $\partial\bar{\psi}(x)$ is
\xl{
	\begin{equation}\label{Eqn:jacobian-n}
	\begin{array}{ll}
	\partial\bar{\psi}(x)&=\hspace{-0.5mm}-2\begin{bmatrix}
	\psi_1(x)\cdot(x-c_1)& \cdots \psi_{n+1}(x)\cdot(x-c_{n+1})
	\end{bmatrix}^{\top}\\
	
	&=\hspace{-0.5mm}-2{\rm{diag}}\{\bar{\psi}(x)\}\begin{bmatrix}
	(x-c_1)& \cdots (x-c_{n+1})
	\end{bmatrix}^{\top},
	\end{array}
	\end{equation}}
where ${\rm{diag}}\{\bar{\psi}(x)\}={\rm{diag}}\{{\psi}_1(x),\cdots,{\psi}_{n+1}(x)\}$.
As ${\rm{diag}}\{\bar{\psi}(x)\}$ is full rank in view of the property of $\psi_i(x)$, for any $x\in\mathcal{X}$,  one has
\begin{equation}\begin{array}{ll}
\text{rank}(\partial\bar{\psi}(x))&=\text{rank}(\begin{bmatrix}
(x-c_1)& \cdots (x-c_{n+1})
\end{bmatrix}^{\top})\\&= n,
\end{array}
\end{equation} 
if and only if $c_1,\cdots,c_{n+1}$ are linearly independent, where the worst testing scenario is $x=c_i$, $i\in\mathbb{N}_{1}^{n+1}$.  \hfill$\square$
\begin{remark}
	Proposition~\ref{Eqn:multiple-choice} implies that, with prescribed Gaussian kernels, one can find $\sum_{i=n}^{n_{\psi}}\frac{n_{\psi}!}{(n_{\psi}-i)!i!}$ combinations of basis functions (choices of $\Psi^{-1}$). In this peculiar case, determining $x$ using multiple $\psi_i'$s, can be stated as a feasibility problem with multiple quadratic equality constraints, while its dual problem does not fulfill the Slater condition that ensures strong duality property, see~\cite{boyd2004convex}.
\end{remark}
}
We recall that (cf.~\cite{korda2018linear}), a straightforward and well-performed choice of $\Psi(x)$ can be of type $\Psi(x)=(x,\bar \Psi(x))$, i.e., with the original state $x$ being included. With this choice, one can promptly find a candidate $\Psi^{-1}(s)=[I_n\  0]s$ such that $\Psi^{-1}(\Psi(x))=x$.

Now, it is possible to state the boundedness property of sets $\bar{\mathcal{W}}$ (i.e., $\mathcal{W}$) and $\mathcal{V}$ in the following proposition.
\xl{
\begin{proposition}\label{prop:uncertainty-bound}
	If $\Psi(x)$ is such that set $\mathcal{S}_{\Psi}$ is bounded, then $\bar{\mathcal{W}}$ and $\mathcal{V}$ are bounded.
\end{proposition}
\textbf{Proof.} To first prove $\bar{\mathcal{W}}$ is bounded, in view of~\eqref{Eqn:non-model} and~\eqref{Eqn:linear_p-residual} and recalling that $s=\Psi(x)$, one has that the modeling uncertainty is $w(s,u,w_o,\hat w_{o})=\Psi(f(\Psi^{-1}(s),u)+w_o)-As-Bu-D\hat w_o$. Letting $\delta w=w_o-\hat w_o$, for all $s\in\mathcal{S}_{\Psi}$, %
$u\in\mathcal{U}$, $\hat w_{o}\in\hat{\mathcal{W}}_o$, and $w_{o}\in\mathcal{W}_o$, it is convenient to write the following inequality:
\begin{equation}\label{Eqn:lipsch-1}
\begin{array}{lll}
\|w(s,u,w_{o},\hat w_{o})\|%
&\leq& L_s \|s\|+
L_u \|u\|+L_{\delta w} \|\delta w_{o}\|+L_{\hat w}\|\hat w_{o}\| \\
&<&+\infty,
\end{array}
\end{equation}
where $L_s$, $L_u$, $L_{\delta w}$, and $L_{\hat w}$ are bounded Lipschitz constants and the last inequality holds since $\mathcal{X}$, $\mathcal{U}$, $\hat {\mathcal{W}}_o$, and $\mathcal{W}_o$ are bounded. Hence, $\mathcal{W}$ is bounded, leading to $\bar{\mathcal{W}}$ being bounded. As for $\mathcal{V}$, it follows from~\eqref{Eqn:linear_p-residual} that $\mathcal{V}=\mathcal{X}\ominus C\mathcal{S}_{\Psi}$ being bounded. 
\hfill $\square$\\
Indeed, the boundedness (instead of convergence) of $\bar w$ and $v$ from Proposition~\ref{prop:uncertainty-bound}   is sufficient for theoretical guarantees of the proposed r-KMPC, which allows us to reduce the dimension of the approximated Koopman model in~\eqref{Eqn:linear_p-residual}. It can also be observed through~\eqref{Eqn:lipsch-1} that, the range of $\bar w$ (i.e., $w$) and $v$ hinges upon the Lipschitz constants $L_s$, $L_u$, and $L_{\hat w}$.
To further reduce the size of $\mathcal{W}$ and $\mathcal{V}$,} it is convenient to minimize $L_s$, $L_u$, and $L_{\hat w}$ or their estimations in the modeling phase by modifying the optimization problem~\eqref{Eqn:appro_K} as 
\begin{equation}\label{Eqn:appro_K-modify}
\min{[\mathcal{K}_{\scriptscriptstyle N_{\phi}}]_{\scriptscriptstyle 1:n_{\psi}},L_s,L_u,L_{\hat w}} V_{\mathcal{K}}+\alpha_sL_s+\alpha_uL_u+\alpha_wL_{\hat w} 
\end{equation}
subject to $L_s,L_u,L_{\hat w}\geq 0$ and 
\begin{equation*}
\|Y_i-Y_j\|\leq L_s \|s_i-s_j\|+L_u \|u_i-u_j\|+ L_{\hat w} \|\hat w_{o,i}-\hat w_{o,j}\|,
\end{equation*}
$i,j\in\mathbb{N}_{1}^{\scriptscriptstyle M}$, where $Y_i=[\mathcal{K}_{\scriptscriptstyle N_{\phi}}]_{\scriptscriptstyle 1:n_{\psi}}(
\Psi(x_i),
\hat {u}_{w,i})-\Psi(x_i^+)$, $\alpha_s,\alpha_u,\alpha_w>0$ are tuning parameters.\\
%
In the case that $w_o=0$, one also has $\hat w_o=0$. Letting	$s_r=\Psi(0)$ and $\bar A=\begin{bmatrix}
I-A^{\top}&C^{\top}
\end{bmatrix}^{\top}$, the triple $(u,s,x)=(0,s_r,0)$ is an equilibrium of model~\eqref{Eqn:linear_p-residual} if and only if 
\begin{equation}\label{Eqn:equili}
\left(w(s_r,0,0,0),v(s_r)\right)=\bar A s_r, 
\end{equation}
which results in a prior knowledge of  $w$ and $v$, 
since $s_r=\Psi(0)$ is available. 
In principle, it is convenient to define an MPC prediction model using~\eqref{Eqn:linear_p-residual}  with $\bar w$ replaced by  $w(s_r,0,0,0)$. 
However, for the case that $w_o$ is time-varying,  
the computation of $w(s_r,0,0,0)$ can be nontrivial because $w_o,\hat w_o=0$ might not be accessible or even exist. As a consequence,~\eqref{Eqn:linear_p-residual} with $\bar w=w(s_r,0,0,0)$ is no longer suitable for prediction in the MPC algorithm. To solve this issue, we first introduce the following proposition.
\begin{proposition}[Equilibrium]\label{prop:1} 
$\left(w(s_r,0,0,0),v(s_r)\right)=0$ if and only if
\begin{equation}\label{Eqn:equili-uncer=0}
s_r\subseteq {\rm{Ker}}\,\bar A \quad \text{or}\quad s_r=0; 
\end{equation}
{\color{black}
	moreover, if the pair $(A,C)$ is observable, condition~\eqref{Eqn:equili-uncer=0} is reduced to
	\begin{equation}\label{Eqn:equili-uncer=0-reduce}
	\quad s_r=0. 
	\end{equation}}
\end{proposition}
\textbf{Proof.} Condition~\eqref{Eqn:equili-uncer=0}  follows directly from~\eqref{Eqn:equili} by linear algebras. 
%
	\begin{remark}
		Condition~\eqref{Eqn:equili-uncer=0} implies that two choices can be chosen for guaranteeing $w(s_r,0,0,0),v(s_r)=0$. A choice is to allow $s_r\neq 0$, i.e., to enforce $s_r\subseteq {\rm Ker}\,\bar A$, which can be imposed by $s_r=As_r$ and $C s_r=0$. The derivation of the above condition requires to 
		include in the optimization problem~\eqref{Eqn:appro_K} the  constraint $As_r=s_r$ and  in~\eqref{Eqn:appro_C} the constraint $Cs_r=0$. The resulting problems can be solved since $s_r$, i.e., $\Psi(0)$ can be chosen a priori.  Note however that the resulting model is not observable if there exist eigenvalues of $A$ being 1, in view of the Popov-Belevitch-Hautus (PBH) test.\\
		Also, provided that a prior condition that $(A,C)$ is observable,~\eqref{Eqn:equili-uncer=0} is equivalent to
		
		\begin{equation}\label{Eqn:linear-obser}
		\begin{array}{l}
		Cs_r=CAs_r=\cdots=
		CA^{n_{\psi}-1}s_r=0
		\end{array}
		\end{equation}
	since $s_r=As_r=A^2s_r,\cdots$ from  $s_r=As_r$,
	leading to the unique solution $s_r=0$.  \\
	Hence, a simple and elegant way is to choose $s_r=0$ for verifying~\eqref{Eqn:equili} (see~\eqref{Eqn:equili-uncer=0-reduce}), which allows to avoid the reformulation of problems~\eqref{Eqn:appro_K} (or~\eqref{Eqn:appro_K-modify}), and~\eqref{Eqn:appro_C}. \hfill $\square$\\
		\end{remark}
	We now give the following basic assumption. }
\begin{assumption}\label{assu:phi-ori}
	The lifted function is constructed such that  $\Psi(0)= 0$.
\end{assumption}
{\color{black}
\begin{remark}
	Assumption~\ref{assu:phi-ori} can be easily verified since, for $\Psi(x)$ being selected as a vector of basis functions with $s_r\neq0$, a coordinate transformation can be used to define a new function $\Psi(x)=\Psi'(x)-\Psi'(0)$, leading to $\Psi(0)=0$, consequently to verification of~\eqref{Eqn:equili}. Take the prescribed Gaussian kernels as an example. Let $\psi'_i(x)=e^{-\|x-c_i\|^2}$, $i\in\mathbb{N}_{1}^{n_{\psi}}$, then it holds that $\psi_i(x)=\psi'_i(x)-\psi'_i(0)=e^{-\|x-c_i\|^2}-e^{-\|0-c_i\|^2}$ with $\psi_i(0)=0$. Hence the resulting $\Psi(x)$ satisfies $\Psi(0)=0$. With this choice, Proposition~\ref{Eqn:multiple-choice} can still be verified because $	\partial\bar{\psi}(x)$  becomes $
	\partial\bar{\psi}(x)=-2{\rm diag}\{\bar{\psi}'(x)\}[(x-c_1)\, \cdots\, (x-c_{n+1})]^{\top}$, where $\bar{\psi}'(x)=(\psi'_1(x),\cdots,\psi'_{n+1}(x))$.
\end{remark}}
{\color{black}	
	\begin{proposition}[Local stabilizability]\label{Eqn:sta-pro}
		Under Assumption~\ref{assu:phi-ori}, model~\eqref{Eqn:linear_p-residual} with $w_o=0$ is stabilizable on the set $\mathcal{S}_{\Psi}\times \mathcal{U}$,  if and only if $f(x,u)$ is stabilizable on the domain $\mathcal{X}\times\mathcal{U}$.
	\end{proposition}

	\textbf{Proof}. (i) Sufficient condition: as $f(x,u)$ is stabilizable on $\mathcal{X}\times\mathcal{U}$, there exists a  feedback law $ u(x)\in\mathcal{U}$ such that 
	$x=0$ is asymptotically stable, under model $x^+=f(x,u)$. In view of the fact that~\eqref{Eqn:linear_p-residual} with $w_o=0$  is equivalent to $x^+=f(x,u)$, it holds that $s=\Psi(x)$ converges to the origin asymptotically under $u(x)$, since $x$ converges to the origin asymptotically and $\Psi(0)=0$. Hence, model~\eqref{Eqn:linear_p-residual} is stabilizable on $\mathcal{S}_{\Psi}\times \mathcal{U}$. \\
	(ii) Necessary condition: likewise, starting with the stabilizability of model~\eqref{Eqn:linear_p-residual} on $\mathcal{S}_{\Psi}\times \mathcal{U}$, one has $s$ converging to the origin asymptotically under $ u(x)\in\mathcal{U}$. The stabilizability of $f(x,u)$ on $\mathcal{X}\times\mathcal{U}$ follows since $x=Cs+v(s)\rightarrow 0+v(0)=0$ asymptotically.
	\hfill $\square$
		\begin{remark}
			Proposition~\ref{Eqn:sta-pro} implies that, in the unperturbed case, i.e., $w_o=0$, there exists a control law such that the point-wise convergence of the closed-loop evolution of $s$ and $x$ under~\eqref{Eqn:linear_p-residual} can be guaranteed, which is a fundamental hint for the result stated in Theorem~\ref{the:rmpc-convergence} (deferred in Section~\ref{sec:theoretical}).
		\end{remark}
	 
In a class of MPC problems such as~\eqref{Eqn:optimiz0} where the penalization on the output is used in the cost function, the observability property could be a prior condition in deriving the closed-loop theoretical property.	 In the following, we also analyze the observability property of~\eqref{Eqn:linear_p-residual}.
	\begin{definition}[Local observability~\cite{albertini1996remarks}]\label{def:observa}
		System~\eqref{Eqn:linear_p-residual} under $w_o=0$ is locally observable on $\mathcal{S}_{\Psi}$ if, for each $s_{a,0},s_{b,0}\in\mathcal{S}_{\Psi}$, the corresponding future states $x_{a,j}\in\mathcal{X}$, $x_{b,j}\in\mathcal{X}$, under the same control sequence, $u_0,\cdots,u_{j}\in\mathcal{U}^{j+1}$, $j\geq 1$, are such that  $x_{a,j}=x_{b,j}$ implies $s_{a,0}=s_{b,0}$.
	\end{definition}
	\begin{proposition}[Local observability]\label{Eqn:obs-pro}
	System~\eqref{Eqn:linear_p-residual} is locally observable on $\mathcal{S}_{\Psi}$ if 
		\begin{subequations}
			{\color{black}
			\begin{align}\label{Eqn:full-map}
			\frac{\partial \{f(x_j,u_j)\circ\cdots\circ f(x_0,u_0)\}}{\partial x_0}=n,
			\end{align}}
for each $x_0\in\mathcal{X}$, $u_i\in\mathcal{U}$, $\forall  i\in\mathbb{N}_0^j$ satisfying $f(x_i,u_i)\in\mathcal{X}$, $\forall  i\in\mathbb{N}_0^j$, and
			\begin{align}\label{Eqn:indpen-basis}
			\Psi(x_a)\neq\Psi(x_b), \forall\, x_a\neq x_b,\ x_a,x_b\in\mathcal{X}.
			\end{align}
		\end{subequations}
	\end{proposition}
	\textbf{Proof}. In view of the argument in~\cite{albertini1996remarks}, if condition~\eqref{Eqn:full-map} is satisfied, it promptly holds that $x_{a,j}=x_{b,j}$ implies $x_{a,0}=x_{b,0}$. Also, $x_{a,0}=x_{b,0}$ implies $\Psi(x_{a,0})=\Psi(x_{b,0})$ since~\eqref{Eqn:indpen-basis} holds.  Finally, $x_{a,j}=x_{b,j}$ implies $s_{a,0}=s_{b,0}$ due to $s_{a,0}=\Psi(x_{a,0})$ and $s_{b,0}=\Psi(x_{b,0})$. Hence,  in view of Definition~\ref{def:observa}, \eqref{Eqn:linear_p-residual} under $w_o=0$ is locally observable on $\mathcal{S}_{\Psi}$. \hfill $\square$ 
	In view of Proposition~\ref{prop:1} and Assumption~\ref{assu:phi-ori},~\eqref{Eqn:unpert} becomes the nominal model of~\eqref{Eqn:linear_p-residual} and is used as the predictor in the online MPC  deferred in Section~\ref{sec:r-KMPC}. 
It is also observed that the pair $(\hat u,\hat s,\hat x)=(0,0,0)$ is an equilibrium of model~\eqref{Eqn:unpert}. 
{\color{black}
	\begin{remark}
		One can conclude from Proposition~\ref{Eqn:sta-pro} and~\ref{Eqn:obs-pro} that  
		$(A,B)$ is stabilizable and $(A,C)$ is observable, if the infimum of Lipschitz constants $ L_s, L_u $, i.e., $\underline L_s,\underline L_u\rightarrow0$ (which leads to $w,v\rightarrow 0$), since model~\eqref{Eqn:linear_p-residual} under $w_o=0$ is equivalent to~\eqref{Eqn:unpert} in this peculiar case. 
		If $\underline L_s,\underline L_u$ are nonzeros, Assumption~\ref{assum:stabi-obser} might not be directly derived using Proposition~\ref{Eqn:sta-pro} and~\ref{Eqn:obs-pro}. For instance, consider a special Koopman model of type: 
		\begin{equation*}\label{Eqn:instable}
		 \left\{\begin{array}{l}
	s_{k+1}={\rm diag}\{1.01,1\}{s}_k+[0\ 1]^{\top} u_k+w_k\\
 x_k=[0\ 1]s_k+v_k,
		\end{array}\right. \qquad
		\end{equation*} 
		where $w=-0.02s$, $v=[0.01\ 0]s$;
which is stabilizable and observable. However, its nominal model is neither stabilizable nor observable.	
	\end{remark}
Hence, we also require the following assumption about~\eqref{Eqn:unpert}.
\xl{
	\begin{assumption}\label{assum:stabi-obser}
		The pair $(A,B)$ is stabilizable and the pair $(A,C)$ is observable.
\end{assumption}} 
Note that, once~\eqref{Eqn:appro_K} and~\eqref{Eqn:appro_C} are solved, one can verify the above assumption via calculating the stabilizability and observability matrices using the computed $A$, $B$, and $C$.\\
}
%
%
\subsection{Robust Koopman MPC design}\label{sec:r-KMPC}
Inline with classic tube-based MPC, the proposed control action to Koopman model~\eqref{Eqn:linear_p-residual} relies on the feedback term of the state error correction:
\begin{equation}\label{Eqn:real_u}
u=\hat u+K(s-\hat s),
\end{equation}
where $ K$ is a gain matrix such that $F=A+B K$ is Schur stable, $\hat u$ and $\hat s$ are decision variables computed with a standard MPC (deferred in~\eqref{Eqn:optimiz1}) with respect to~\eqref{Eqn:unpert}. 
{\color{black}
	\begin{remark}
		Different from classic tube-based MPC for linear systems~\cite{mayne2005robust}, $\hat u$ and $K(\Psi(x)-\hat s)$ in~\eqref{Eqn:real_u}  are both nonlinear control laws on the original state $x$ since $\hat u$ depends on $\hat s$ and $\hat s$ is related to $s=\Psi(x)$, see~\eqref{Eqn:inin-condi}.	 
	\end{remark}
} 
From~\eqref{Eqn:linear_p-residual} and~\eqref{Eqn:real_u}, the error $e_{\scriptscriptstyle {s}}:={s}-\hat {s}$ evolves according to the following unforced system \eqref{Eqn:linear_p-residual}:
\begin{equation}\label{Eqn:e}
 \left\{\begin{array}{l}
e_{\scriptscriptstyle{s}}^+=Fe_{\scriptscriptstyle{s}}+\bar w\\
[0.2cm] e_x=C e_{\scriptscriptstyle{s}}+v,
\end{array}\right. \qquad
\end{equation}
where $e_x=x-\hat x$.
Let $\mathcal{Z}_s$ be a robust positively invariant set of $e_s$ such that $\mathcal{Z}_s\subseteq F\mathcal{Z}_s\oplus \bar{\mathcal{W}}$, then it holds that $e_x\in C\mathcal{Z}_s\oplus \mathcal{V}:=\mathcal{Z}_x$.\\
Now we are ready to state the nominal MPC problem to compute $\hat u$ and $\hat s$ in~\eqref{Eqn:real_u}.
At any time instant $k$, the following online quadratic optimization problem is to be solved:
\begin{equation}\label{Eqn:optimiz1}
\min{\hat s_k,{\bm {\hat u}}_{k}}  {V\big(\hat{s}_k,\bm {\hat u}_{\scriptscriptstyle k}\big)}
\end{equation}
where 
\begin{equation}\label{Eqn:V_b}
\begin{array}{ll}
V\big(\hat{s}_k,\bm {\hat u}_{\scriptscriptstyle k}\big)=
\sum_{i=0}^{N-1}(\|\hat {x}_{k+i}\|_{ Q}^2+\|\hat u_{k+i}\|_{R}^2)
+V_f(\hat{s}_{k+N}),
\end{array}
\end{equation}
and $\hat s_k$, ${\bm {\hat u}}_{\scriptscriptstyle k}$ are the decision variables, \xl{$V_f(\hat{s})$ is the terminal cost with respect to the nominal lifted state $\hat s$ and it is chosen as $V_f(\hat{s})=\hat{s}^{\top}P\hat{s}$,} where the symmetric positive-definite matrix $P$ is the solution to the Lyapunov equation
\begin{equation}\label{Eqn:LYA}
F^{\top}PF-P=-(\bar Q+K^{\top}RK)
\end{equation}
and $\bar Q=C^{\top}QC$.\\
The optimization problem~\eqref{Eqn:optimiz1} with~\eqref{Eqn:V_b} is performed subject to the following constraints:
\begin{enumerate}[1)]
	\item The nominal Koopman model~\eqref{Eqn:unpert}.
	\item Tighter state and control constraints:
	\begin{subequations}\label{Eqn:constraint-MPC}
		\begin{align}
		\hat {s}_{k+i}\in   {\mathcal{S}},\,\,i=0,\dots,N-1\\
		\hat u_{k+i}\in \hat{ {\mathcal{U}}},\,\,i=0,\dots,N-1, 
		\end{align}
		where   ${\mathcal{S}}=\{\hat{s}|C\hat s\in  {\mathcal{X}}\ominus{\mathcal{Z}}_x\}$, ${\hat {\mathcal{U}}}={\mathcal{U}}\ominus K{\mathcal{Z}}_s$.
		\item The initial and terminal state constraints 
		\begin{align}
			s_k-\hat {s}_k\in \mathcal{Z}_s\label{Eqn:inin-condi}\\
		  \hat {s}_{k+N}\in {{\mathcal{S}}_f},
		\end{align}
	\end{subequations}
\end{enumerate}
where ${\mathcal{S}}_f$ is a positive invariant set of~\eqref{Eqn:unpert} under state and control constraints such that $(A+BK){\mathcal{S}}_f\subseteq {\mathcal{S}}_f$. \\
Assume that, at any time instant $k$ the optimal solution of $\hat{s}_{k}$ and $\bm {\hat u}^{\ast}_{k}$ can be found and is denoted by  $\hat{s}_{k|k}$ and $\bm {\hat u}^{\ast}_{k|k}$, then the final control applied to system~\eqref{Eqn:non-model} is given as
\begin{equation}\label{Eqn:real_u_k}
u_k=\hat u^{\ast}_{k|k}+K(s_k-\hat{s}_{k|k}).
\end{equation}
In summary, the pseudocode of the proposed r-KMPC is described in Algorithm~\ref{alg:r-KMPC}. 
\begin{algorithm}[h]
	\caption{Pseudocode of r-KMPC}
	\label{alg:r-KMPC}
	\textbf{Off-line designs:}
	\begin{algorithmic}
		\State \textbf{1)} Select $\Psi(x)$ such that Assumption~\ref{assum:invers-phi} is verified.
		\State \textbf{2)} Compute $A$, $B$, $C$, and $D$ with~\eqref{Eqn:appro_K} (or~\eqref{Eqn:appro_K-modify}) and~\eqref{Eqn:appro_C}, and check that Assumption~\ref{assum:stabi-obser} is verified.
		\xl{\State \textbf{3)} Calculate $\bar{\mathcal{W}}$ and $\mathcal{V}$ according to Algorithm~\ref{alg:com-sets} described in Appendix~\ref{sec:app-b}.}
		\State \textbf{4)} Compute $P$ and $K$ with~\eqref{Eqn:LYA} and calculate the robust positively invariant set $\mathcal{Z}_s$, $\mathcal{Z}_x$ with $K$.
		\State \textbf{5)} Compute  $\mathcal{S}$, $\hat {\mathcal{U}}$ and check that Assumption~\ref{assump:nonemptyset} is verified (see Section~\ref{sec:theoretical}); calculate the terminal set $\mathcal{S}_f$ with $K$.
	\end{algorithmic}
	\dotfill \\
	\textbf{On-line procedures:}\\
	At each discrete-time step $k=1,2,\cdots$
	\begin{algorithmic}
		\State \textbf{1)} Measure $x_k$ and set the lifted state $s_k=\Psi(x_k)$.
		\State \textbf{2)} Solve~\eqref{Eqn:optimiz1} with~\eqref{Eqn:V_b} and obtain ${\bm\hat u}^{\ast}_{k|k}$, $\hat s_{k|k}$.
		\State \textbf{3)} Set $u_k$ with~\eqref{Eqn:real_u_k} and apply it to the nonlinear system~\eqref{Eqn:non-model}.
	\end{algorithmic}
\end{algorithm}
\subsection{Closed-loop property analysis}\label{sec:theoretical}
\xl{
\begin{assumption}\label{assump:nonemptyset}
	The computed nominal control and lifted state constraints are non-empty and contain the origin in their interiors, i.e., $\{\bm 0\}\subset\hat{\mathcal{U}}$, $\{\bm 0\}\subset{\mathcal{S}}$.
\end{assumption}
Under Assumptions~\ref{Eqn:data-distribution} and~\ref{assum:invers-phi}, to satisfy the  above condition, one can choose a proper observable function $\Psi(x)$ and a proper number of samples $M$ used in~\eqref{Eqn:appro_K} and~\eqref{Eqn:appro_C}.}
\begin{theorem}[Recursive feasibility]\label{the:recur-rmpc}
	\xl{Under Assumptions \ref{Eqn:data-distribution}-\ref{assump:nonemptyset}}, if the optimization problem~\eqref{Eqn:optimiz1} is feasible at the initial time $k=0$, then it is recursively feasible at all times $k\in\mathbb{N}_1^{\infty}$, i.e., the recursive feasibility of r-KMPC is guaranteed.
\end{theorem}
\textbf{Proof.} Assume at any time instant $k$, an optimal decision sequence of~\eqref{Eqn:optimiz1} with~\eqref{Eqn:V_b}, i.e., $\hat s_{k|k}$, $\bm {\hat u}^{\ast}_{k|k}$, have been found 
associated with the optimal cost, denoted as $V^{\ast}_k$, such that the state and control constraints $\hat{s}_{k+i}\in {\mathcal{S}}$,  $\hat {u}_{k+i}\in{\hat{\mathcal{U}}}$, $\forall i\in\mathbb{N}_0^{N-1}$, and the terminal constraint $\hat {s}_{k+N|k}\in{\mathcal{S}}_f$ is fulfilled. Hence, it holds that the real state and control constraints are verified, i.e., $x_{t|k}\in\mathcal{X}$ and $u_{t|k}\in\mathcal{U}$ for all $t\geq k$. At the next time instant $k+1$, choose $\hat{s}_{k+1|k+1}=\hat{s}_{k+1|k}$, $\bm{\hat u}^s_{k+1}=\bm {\hat u}^{\ast}_{k+1:k+N-1|k}, K\hat{s}_{k+N|k}$ as a candidate sub-optimal solution, under which it follows that $C{s}_{k+1}+v_{k+1}\in\mathcal{X}$, $\hat {s}_{k+1|k}\in   {\mathcal{S}}$, leading to ${s}_{k+1}-\hat{s}_{k+1|k}\in {\mathcal{Z}}_s,$ $ 
\hat s_{k+j|k+1}\in {\mathcal{S}},\, \forall j\in\mathbb{N}_{1}^{N},$
 through  inheritance,
and $\hat{s}_{k+N+1|k+1}\in{\mathcal{S}}_f$ in view of the definition of ${\mathcal{S}}_f$.  
Hence,~\eqref{Eqn:optimiz1} is feasible at time $k+1$ under $\hat s_{k+1|k}$, $\bm{\hat u}^s_{k+1}$, associated with a sub-optimal cost $V^s_{k+1}$. The recursive feasibility of  r-KMPC holds. \hfill$\square$\\
Under Theorem~\ref{the:recur-rmpc}, one can state the following theoretical results.
\begin{theorem}[Closed-loop robustness]\label{the:rmpc-robust} \xl{Under Assumptions \ref{Eqn:data-distribution}-\ref{assump:nonemptyset}, it holds that:} 
	\begin{enumerate}[(a)] 
		
		\item 	the lifted nominal system~\eqref{Eqn:unpert} using $\hat s_k,\hat u_k$ computed with~\eqref{Eqn:optimiz1} converges to the origin asymptotically, i.e., 	$\hat s_k,\, \hat x_k,\,\text{and}\,\hat u_k\rightarrow 0\ \text{as}\ k\rightarrow+\infty;$
		\item the state $s$ and the control $u$ of the closed-loop system~\eqref{Eqn:linear_p-residual} with~\eqref{Eqn:real_u}  are such that
		
		$$s_k\rightarrow{\mathcal{Z}}_s\ \text{and}\ u_k\rightarrow K{\mathcal{Z}}_s\ \text{as}\ k\rightarrow+\infty;$$
		and   $$x_k\rightarrow {\mathcal{Z}}_x\ \text{as}\ k\rightarrow+\infty.$$
	\end{enumerate}
\end{theorem}
\textbf{Proof.} To prove the closed-loop robustness property, first note that the optimal cost $V^{\ast}_k$ associated with the optimal solution $\hat s_{k|k}$, $\bm {\hat u}^{\ast}_{k|k}$ (cf. Theorem~\ref{the:recur-rmpc}), satisfies  $V^{\ast}_k\geq \|\hat {x}_k\|_{ Q}^2$ for all $\hat x_k\oplus\mathcal{Z}_x\in\mathcal{X}$. Also, iterating~\eqref{Eqn:LYA} leads to $V^{\ast}_k\leq \hat s_k^{\top}P\hat s_k$ for $\hat s_k\in{\mathcal{S}}_f$. Considering the fact that $V^{\ast}_{k+1}\leq V^s_{k+1}$, where $V_k^s$ is the sub-optimal cost associated with $\hat s_{k+1|k}$, $\bm{\hat u}^s_{k+1}$ (cf. Theorem~\ref{the:recur-rmpc}), and in view of~\eqref{Eqn:LYA}, one has the monotonic property:
\begin{equation}\label{Eqn:V-MONO-r}
\begin{array}{lll}
V^{\ast}_{k+1}- V^{\ast}_k&\leq&-\|\hat {x}_k\|_{Q}^2-\|\hat u_k\|_{R}^2+\|\hat {s}_{k+N}\|^2_{F^{\top}PF-P+\bar Q+K^{\top}RK}\\
&=&-\|\hat {x}_k\|_{Q}^2-\|\hat u_k\|_{R}^2.
\end{array}
\end{equation}
Then, from~\eqref{Eqn:V-MONO-r},  $V^{\ast}_{k+1}- V^{\ast}_k\rightarrow0$ as $k\rightarrow+\infty$. Consider also $$ V^{\ast}_k-V^{\ast}_{k+1}\geq \|\hat {x}_k\|_{ Q}^2+\|\hat u_k\|_{R}^2,$$ hence $\hat u_k\rightarrow 0$, $\hat x_k\rightarrow 0$, as $k\rightarrow +\infty$, in view of the positive-definiteness of $R$ and $Q$. Also, $\hat{s}_k\rightarrow 0$ as $k\rightarrow +\infty$ due to $\Psi(0)=0$ in view of Assumption~\ref{assu:phi-ori}.\\ 
Moreover, recall that $x\in\hat x\oplus \mathcal{Z}_x$, $u\in\hat u\oplus K\mathcal{Z}_s$, and $s\in\hat s\oplus \mathcal{Z}_s$, it holds that, $$ x_k\rightarrow {\mathcal{Z}}_x\ \text{ and }\ u_k\rightarrow K{\mathcal{Z}}_s\ \text{ as }\ k\rightarrow +\infty,$$ and $$s_k\rightarrow{\mathcal{Z}}_s\ \text{as}\ k\rightarrow +\infty.$$\hfill$\square$
{\color{black}
\begin{remark}
	Note that, in the MPC problem~\eqref{Eqn:optimiz1} with~\eqref{Eqn:V_b}, penalizing the lifted state, i.e., $\|\hat s\|_{\tilde Q}^2$ can be used instead, where $\tilde Q\in\mathbb{R}^{n_{\psi}\times n_{\psi}}$ is a positive-definite tuning matrix. Herewith the observability of $(A,\,C)$ is not necessarily required in~\eqref{Eqn:LYA} since a solution surely exists for any  $\tilde Q$ being positive-definite. In this case, in the proof argument of Theorem~\ref{the:rmpc-robust}, $\|\hat x\|_Q^2$ in~\eqref{Eqn:V-MONO-r}  is replaced by $\|\hat s\|_{\tilde Q}^2$, which  leads directly to the asymptotic stability of $\hat s$ and $\hat x$ since $\hat x=C\hat s$. 
\end{remark}}
Moreover, under the result in Theorem~\ref{the:rmpc-robust}, point-wise asymptotic convergence of r-KMPC  can be verified, under $w_o=0$.
\begin{theorem}[Point-wise convergence]\label{the:rmpc-convergence} \xl{Under Assumptions \ref{Eqn:data-distribution}-\ref{assump:nonemptyset},} if $w_o=0$ and  \begin{equation}\label{Eqn:iss-con}
	E=(L_sI+L_uK^{\top}K)\|\sum_{k=0}^{+\infty}F^k\| \ \text{is Schur stable,}
	\end{equation}
	then the closed-loop systems~\eqref{Eqn:linear_p-residual} with~\eqref{Eqn:real_u} and~\eqref{Eqn:non-model} with~\eqref{Eqn:real_u} converge to the origin asymptotically, i.e., 
	$$x_k\rightarrow 0,\ \ u_k\rightarrow 0,\   \text{and}\ s_k\rightarrow 0\ \text{as}\ k\rightarrow+\infty.$$
	
\end{theorem}
\textbf{Proof.} \xl{In view of Theorem~\ref{the:rmpc-robust}.(a),} assume that $\hat u$, $\hat x$, and $\hat{s}$ have converged asymptotically to the origin, i.e.,   $\hat u=0$, $\hat x=0$, and $\hat{s}=0$.
Also, $u$, $x$, and ${s}$ have converged asymptotically to robust tubes, i.e., $ x\in {\mathcal{Z}}_x,\ u\in K{\mathcal{Z}}_s,$ and $s\in{\mathcal{Z}}_s.$ \\
From~\eqref{Eqn:e}, one can write
\begin{equation}\label{Eqn:x-con}
\left\{\begin{array}{l}
s_{k+1}=Fs_k+w_k\\
[0.2cm] x_k=C s_k+v_k,
\end{array}\right. \qquad
\end{equation}
{\color{black}
	where, in view of~\eqref{Eqn:lipsch-1}, the uncertainty $w$ satisfies: 
	\begin{equation}\label{Eqn:w-con-bound}
	\begin{array}{ll}
	\|w\|&=\|w(s,u,0,0)\|\\
	&\leq L_s\|s\|+L_u\|u\|\\
	&=L_s\|s\|+L_u\|\hat u+K(s-\hat s)\|\\
	&\leq\|s\|_{L_s I+L_uK^{\top}K}
	\end{array}
	\end{equation} and the last inequality is due to~\eqref{Eqn:real_u} and $\hat u,\hat s=0$. 
	Inline with~\cite{farina2018hierarchical}, we write the evolution $s_{k+1}=Fs_k+w_k$ from~\eqref{Eqn:x-con} with two redundant interconnected systems, i.e.,
	\begin{equation}\label{Eqn:x-redunt}
	\left\{\begin{array}{l}
	s_1^+=Fs_1+w_2\\
	s_2^+=Fs_2+w_1,
	\end{array}\right. \qquad
	\end{equation}
	where the initial condition $s_{1,0}=s_{2,0}=s_0$ and $w_{1,0}=w_{2,0}=w_0$.} 
In view of the small gain theorem in~\cite{farina2018hierarchical}, the interconnected system converges to the origin asymptotically~\eqref{Eqn:x-redunt} if 
\begin{equation}
\bar {E}=\begin{bmatrix}
0& E\\
E&0
\end{bmatrix}%
\end{equation}
is Schur stable, which is fulfilled by~\eqref{Eqn:iss-con}. 
Hence, the state $s$ converges to the origin asymptotically. Moreover, $x_k\rightarrow 0\ \text{as}\ k\rightarrow+\infty$ since $x_k=Cs_k+v(s_k)$, $s_k\rightarrow 0$ as $k\rightarrow +\infty$, and $v(0)=0$. Also, in view of~\eqref{Eqn:real_u},  $u_k\rightarrow 0\ \text{as}\ k\rightarrow+\infty$.
\hfill$\square$
\section{Simulation results}\label{sec:simulation}
\subsection{Regulation of a Van der Pol oscillator}\label{sec:example-van}
{\color{black}
	Consider a Van der Pol oscillator~\cite{korda2018linear}, whose continuous-time model is
		\begin{equation}\label{Eqn:ct}
		\begin{array}{ll}
		\begin{bmatrix}
		\dot x_1\\
		\dot x_2
		\end{bmatrix}=&\begin{bmatrix}
		x_2\\
		2x_2-10x_1^2x_2-0.8x_1-u
		\end{bmatrix}+w_o,
		\end{array}
		\end{equation}
	where $x_1$ and $x_2$ represent the position and speed  respectively, while $u$ is the force input,  $\|w_o\|_{\infty}\leq 0.4$. Peculiarly, a sinusoidal noise $w_o=0.4{\rm sin} (10\pi t)$ is initially adopted.\xl{The consideration on other types of disturbances  are deferred in Table~\ref{tab:Tab_com1}.} Denoting $x=(x_1,x_2)$, the state and control are limited as
	$-(2.5m,2.5m/s)\leq x\leq(2.5m,2.5m/s),\ 
	-10m^2/s\leq u\leq 10m^2/s.$\\
	\begin{figure}[h]
		\centering
		\includegraphics[width=0.5\textwidth]{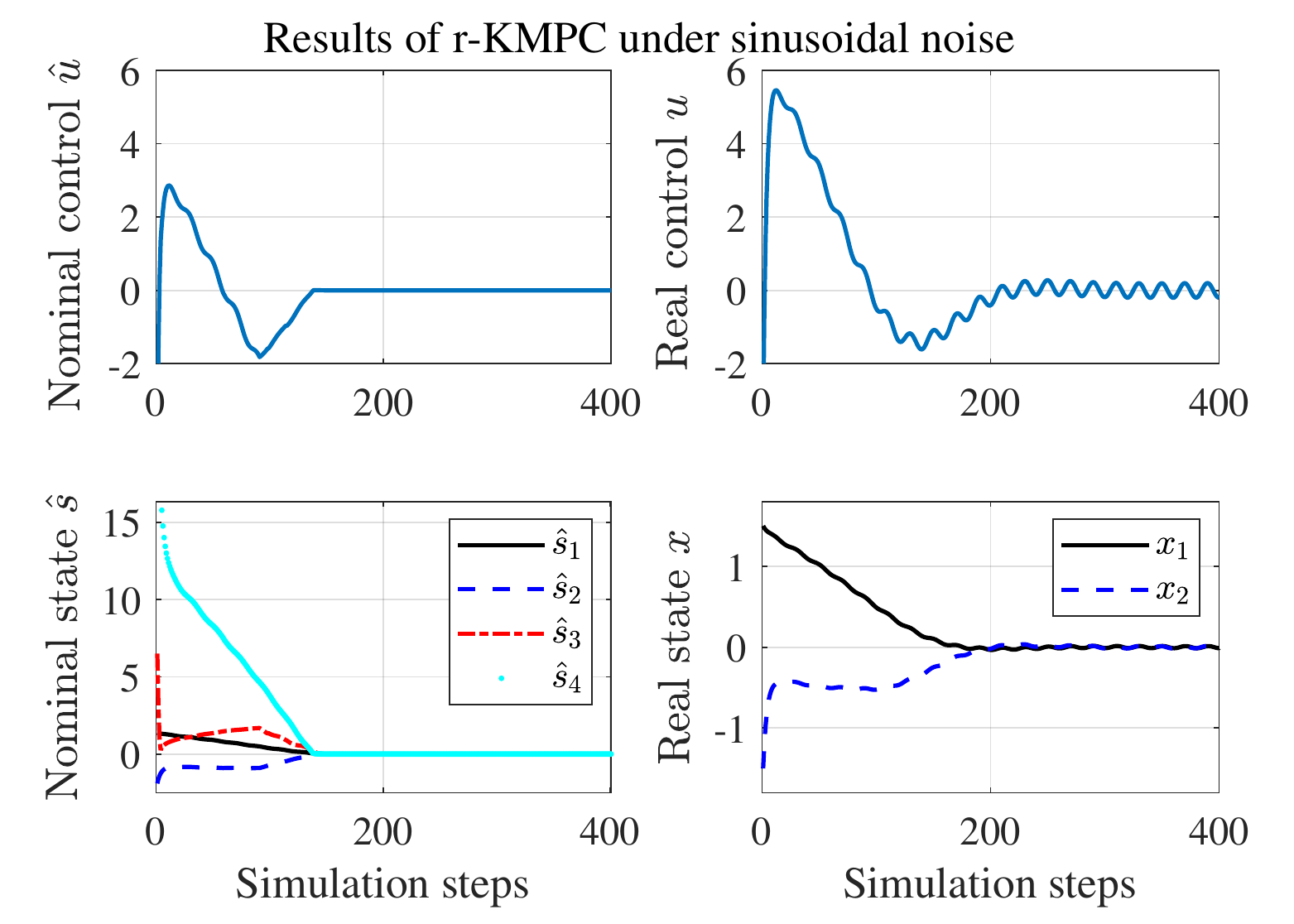} 
		\caption{The  state and control of the controlled Van der Pol oscillator with r-KMPC under sinusoidal noise.}
		\label{fig:sta-con-sin}
	\end{figure}
	\begin{figure}[h]
		\centering
		\includegraphics[width=0.5\textwidth]{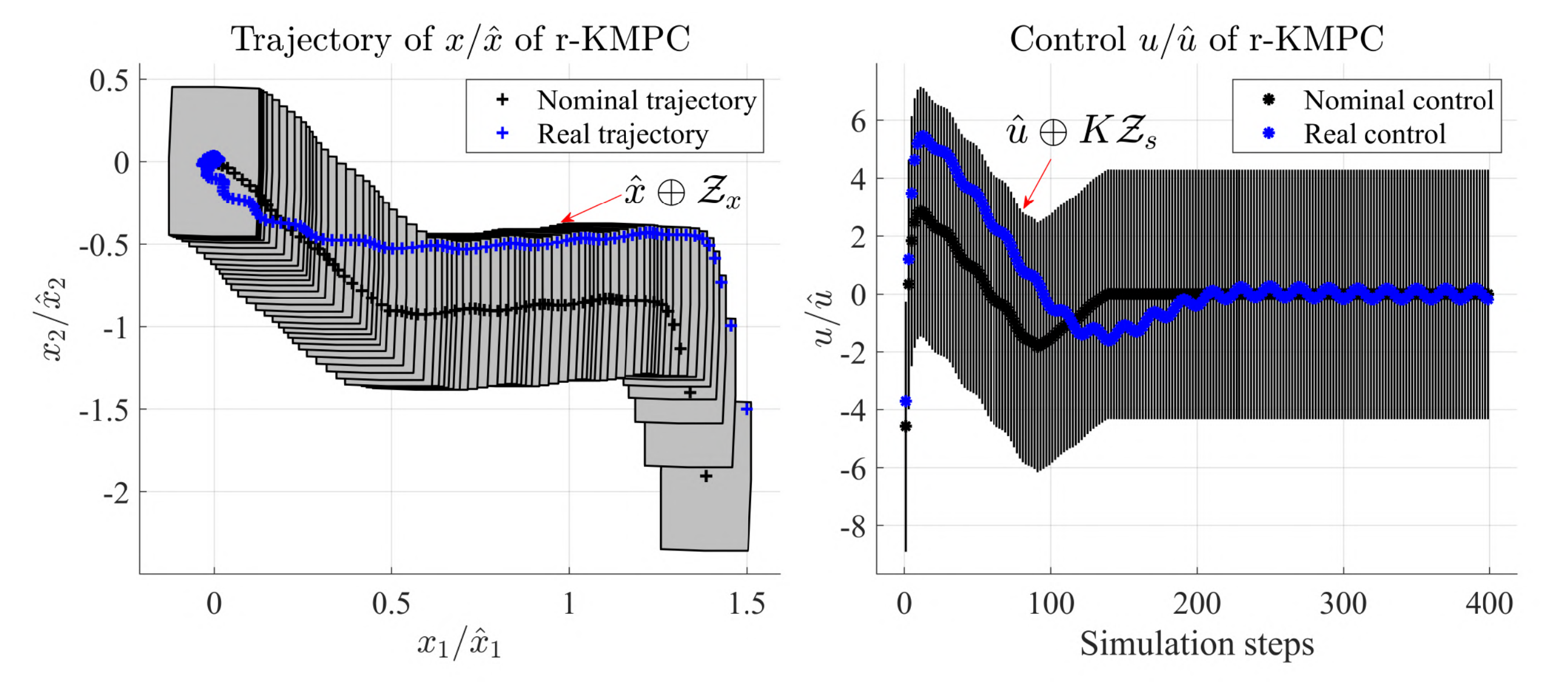}
		\caption{The state and control tubes of the controlled Van der Pol oscillator under sinusoidal noise. In the left (right) panel, the real trajectory, marked with blue + ($\ast$), lies in the grey tube, i.e., $\mathcal{Z}_x$ ($K\mathcal{Z}_s$), centered at the nominal one, marked with black + ($\ast$).}
		\label{fig:real-tra-sin}
	\end{figure}
%
%
	\begin{figure}[h]
		\centering
		\includegraphics[width=0.5\textwidth]{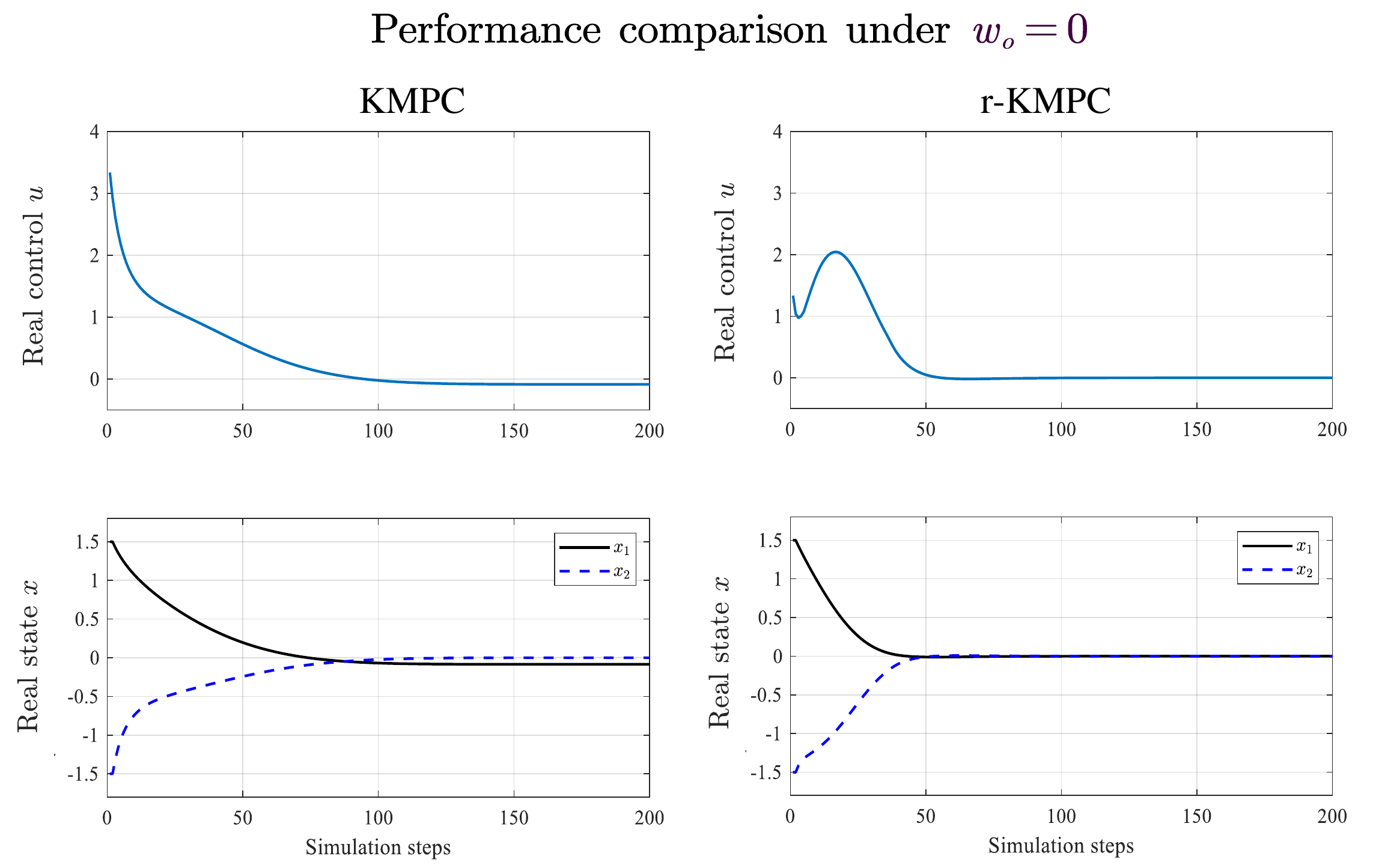} 
		\caption{Performance comparison of the controlled Van der Pol oscillator with r-KMPC and KMPC under $w_o=0$. The state and control of r-KMPC converge asymptotically to the origin.}
		\label{fig:sta_con_no_external_dis}
	\end{figure}
	In order to implement r-KMPC, we first have discretized model~\eqref{Eqn:ct} to obtain model~\eqref{Eqn:non-model} with a suitable sampling period $T=0.01s$. We  have obtained the data-driven model using Koopman operators. The datasets of $(u,\hat w_{o}, x,x^+)$ in~\eqref{Eqn:appro_K} with $M=8\cdot 10^5$ have been collected with a uniformly distributed random (UDR) control policy. 
	A thinplate BF has been selected to construct the lifted observable state variables, i.e., $\Psi(x)=(x,\psi_1(x),\psi_2(x))-(\bm 0,\psi_{1}(0),\psi_{2}(0))$, where  $\psi_i(x)=\|x-c_i\|^2\log(\|x-c_i\|),$ $i=1,2$, $n_{\psi}=4$, $c_1=(0.381,
		-0.341)$, $c_2=(0.267,-0.889)$ are the kernel centers generated with UDR numbers. 
	{\color{black}  
		The parameters of the linear predictor have been computed according to~\eqref{Eqn:appro_K}   and~\eqref{Eqn:appro_C}.\\
		The penalty matrices $\tilde Q$ and $R$ have been selected as $\tilde Q=\text{diag}\{1,1,0.1,0.1\}$, $R=0.1$.   The terminal penalty matrix can be calculated with~\eqref{Eqn:LYA}. 
		The sets $\bar{\mathcal{W}}$, $\mathcal{V}$ have been empirically obtained according to Algorithm 2, see the Appendix. 
		The gain matrix, associated with the computation of $\mathcal{Z}_s$, has been computed as $K=[6.5\	4.3\	-0.4\	0.1]$,  resulting in the closed-loop poles of the linear system $\{0.91,\
		0.93,\ 0.98 + 0.003i,\ 0.98 - 0.003i\}$.
		The robust invariant set has been obtained with~\eqref{Eqn:e}  and the terminal constraint ${\mathcal{S}}_f$ has been computed, according to the algorithms described in~\cite{rawlings2009model}.
		The prediction horizon has been chosen as $N=10$. \\
		The proposed r-KMPC has  been implemented with an initial condition $x_0=(1.5,-1.5)\in \mathcal{X}$. 
		The simulation tests with r-KMPC have been performed within Yalmip toolbox~\cite{Lofberg2004} installed in Matlab 2019a environment. 
		The simulation results  are reported in Figures~\ref{fig:sta-con-sin}-\ref{fig:sta_con_no_external_dis}. It can be seen from Figure~\ref{fig:sta-con-sin} that the nominal lifted state $\hat s$ and control $\hat u$ converge to the origin; and the real state and control converge to the neighbor of the origin, under sinusoidal noises. Note that, the real state trajectory $x$ and the real control $u$ remain in tubes centered at $\hat x$ and $\hat u$ respectively, i.e., $x\in\hat x\oplus \mathcal{Z}_s$ and $u\in\hat u\oplus K\mathcal{Z}_x$, which verifies the robustness of the closed-loop control system, see Figure~\ref{fig:real-tra-sin}. Also, as shown in Figure~\ref{fig:sta_con_no_external_dis}, under no external perturbation, the state and control of r-KMPC converge asymptotically to the origin. \\
		\begin{table}[h!tb]
		\centering \caption{Cumulative cost comparison: ``S" and ``SW" stand for ``sinusoidal" and ``Step-wise" respectively. }
		\label{tab:Tab_com1}
		\vskip 0.2cm
		\scalebox{0.8}{
			\begin{tabular}{ccccccc}
				\hline
				\multicolumn{3}{c}{\multirow{2}{*}{Algorithm}}& r-KMPC &\multicolumn{3}{c}{KMPC~\cite{korda2018linear}}\\
				\cline{5-7}
				\multicolumn{3}{c}{}	& $n_{\psi}=4$ 	& $n_{\psi}=4$  & $n_{\psi}=12$ & $n_{\psi}=22$   \\
				\hline
				\multirow{8}{*}{Cost $J$} 
				&\multirow{4}{*}{Thinplate kernel}&Nominal  & $258$ & $431$ &  $401$&360\\
				&&	S-noise  & $270$ & $449$ &  $418$&374\\
				& &	UDR-noise  & $248$ & $413$ &  $403$&371\\
				& &	SW-noise  & $262$ & $379$ &  $387$&359\\
				\cline{2-7}
				&\multirow{4}{*}{Polynomial kernel}&	Nominal  & $247$ & $427$ &  $402$&--\\
				&&	S-noise  & $257$ & $446$ &  $419$&--\\
				& &	UDR-noise  & $241$ & $416$ &  $403$&--\\
				& &	SW-noise  & $253$ & $416$ &  $410$&--\\
				\hline
			\end{tabular}
		}	 
	\end{table}
		Numerical comparison with KMPC~\cite{korda2018linear}: We have also compared the proposed r-KPMC with the Koopman MPC in~\cite{korda2018linear} in terms of control performance.  First, we have chosen the lifted function in KMPC as $\Psi(x)$ without the resetting procedure, i.e., $\Psi(0)$ is not identically zero. The resulting one-step prediction collective square error with 50000 different initial state conditions is 56.4, comparable with  55.9 obtained by using the resetting procedure. 
		As shown in Figure~\ref{fig:sta_con_no_external_dis}, the resulting controller using KMPC does not yet converge with 400 steps, while the state and control of our approach converge to the origin.  For a numerical comparison, the cumulative costs of both approaches, computed as  $J=\sum_{k=1}^{400}\|x_k\|_Q^2+\|u_k\|^2_R$, are collected in Table~\ref{tab:Tab_com1}, which show that the proposed r-KMPC results in a lower cost consumption, compared with KMPC~\cite{korda2018linear} even using a greater number of observable functions.
		\xl{
	\subsection{Angle regulation of an inverted pendulum}
	Consider the problem of angle regulation of an inverted pendulum, whose continuous-time model is
		\begin{equation}\label{Eqn:ct-in}
	\begin{array}{ll}
	\begin{bmatrix}
	\dot x_1\\
	\dot x_2
	\end{bmatrix}=&\begin{bmatrix}
	x_2\\
	4g{\rm sin}(x_1) -3u{\rm cos}(x_1)
	\end{bmatrix}+w_o,
	\end{array}
	\end{equation}
	where $x=(x_1,x_2)$ are the state variables that represent the angle and its rate respectively, while $u$ is the control, $w_o=2{\rm sin} (10\pi t)$. 
	The constraints to be enforced are
	$-(1 rad,2rad/s)\leq x\leq(1rad,2rad/s),\ 
	-20m^2/s\leq u\leq 20m^2/s.$\vspace{-3mm}\\

	\begin{figure}[h]
		\centering
		\includegraphics[width=0.5\textwidth]{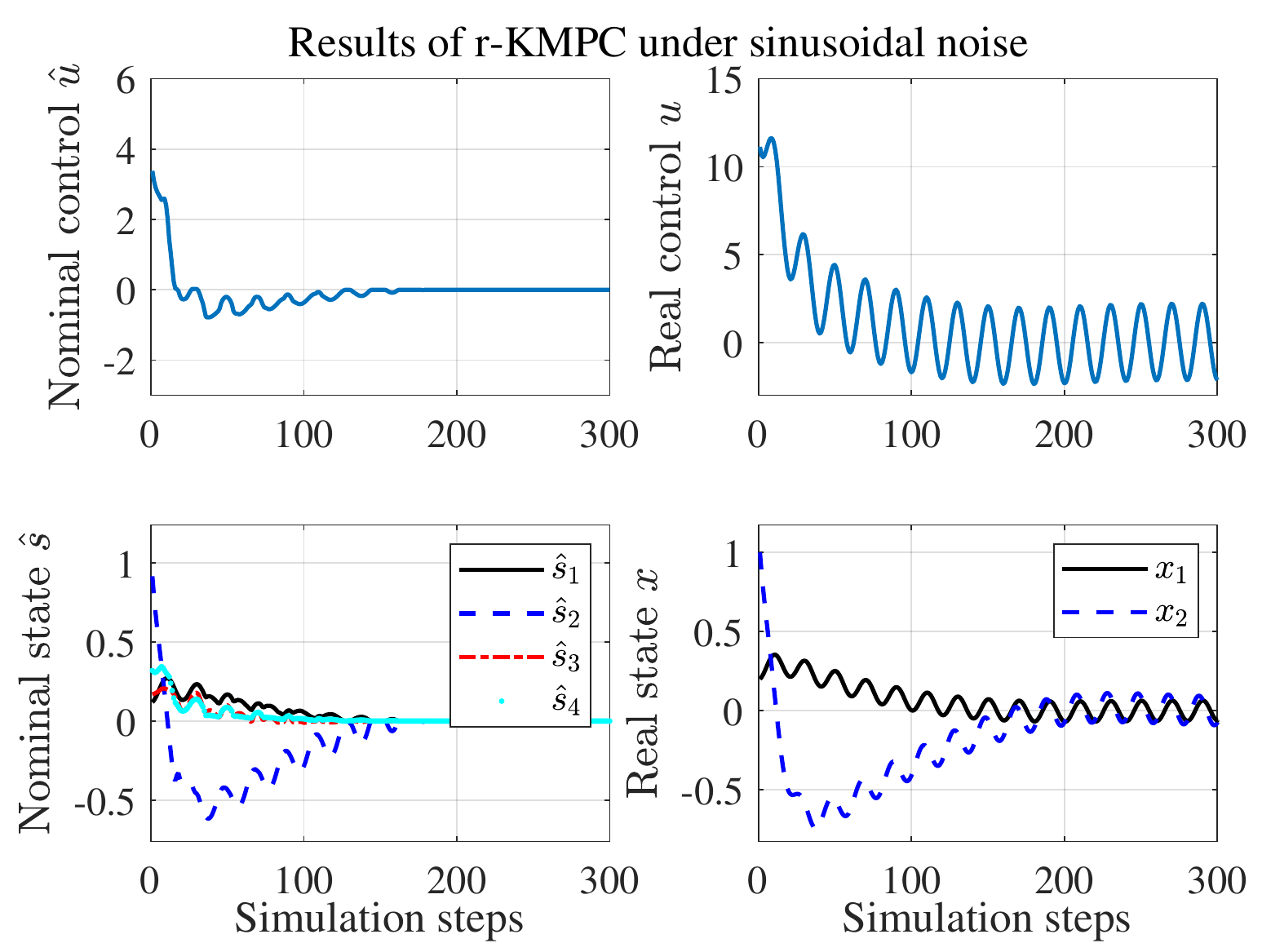} 
		\caption{The  state and control of the controlled inverted pendulum with r-KMPC under sinusoidal noise.}
		\label{fig:sta-con-sin-ip}
	\end{figure}
	\begin{figure}[h]
		\centering
		\includegraphics[width=0.5\textwidth]{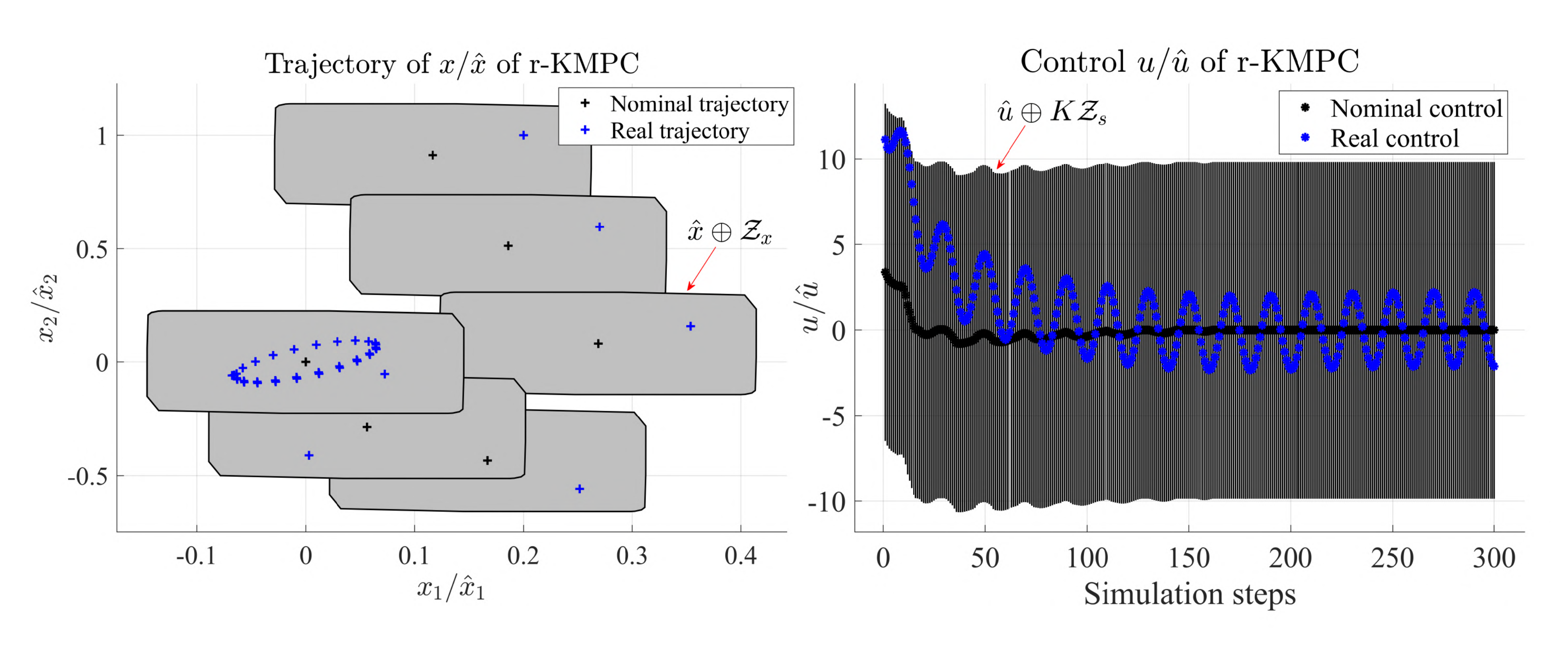}
		\caption{The state and control tubes of the controlled inverted pendulum under sinusoidal noise. In the left (right) panel, the real trajectory, marked with blue + ($\ast$), lies in the grey tube, i.e., $\mathcal{Z}_x$ ($K\mathcal{Z}_s$), centered at the nominal one, marked with black + ($\ast$).}
		\label{fig:real-tra-sin-ip}
	\end{figure}
	%
	%
	\begin{figure}[h]
		\centering
		\includegraphics[width=0.5\textwidth]{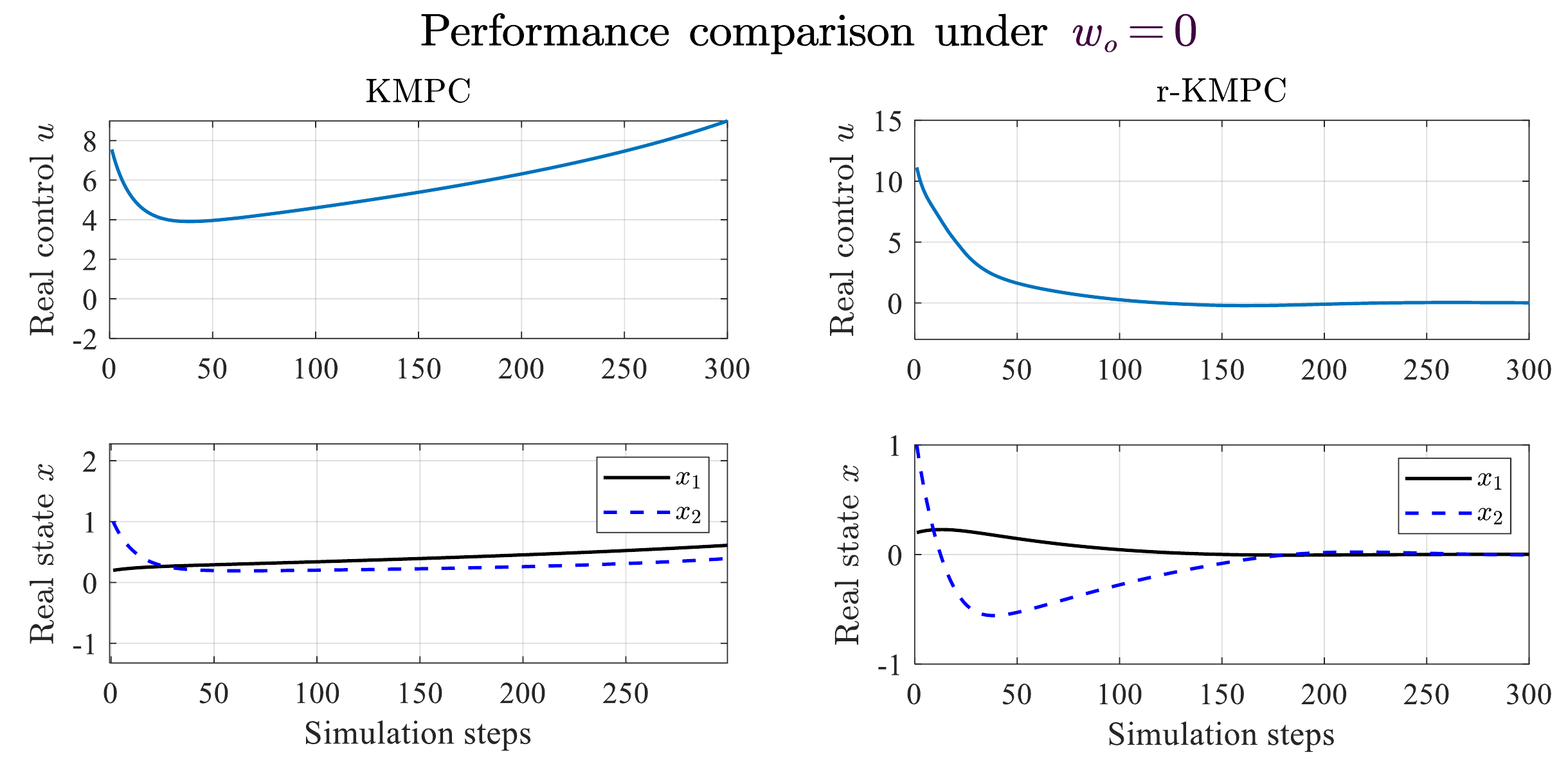} 
		\caption{Performance comparison of of the controlled inverted pendulum with r-KMPC and KMPC under $w_o=0$. The state and control of r-KMPC converge asymptotically to the origin.}
		\label{fig:sta_con_no_external_dis-ip}
	\end{figure}
	In line with Section~\ref{sec:example-van}, model~\eqref{Eqn:ct-in} is discretized with $T=0.005s$. The data samples have been obtained with $M=5\cdot 10^4$ under a UDR control policy.  A Gaussian kernel has been selected to construct $\Psi(x)$, i.e., $\Psi(x)=(x,\psi_1(x),\psi_2(x),\psi_3(x))-(\bm 0,\psi_{1}(0),\psi_{2}(0),\psi_3(0))$, where  $\psi_i(x)=e^{-\|x-c_i\|^2},$ $i=1,2,3$, $n_{\psi}=5$, $c_1=(-0.644,
	-1.09)$, $c_2=(-0.99,0.76)$, and $c_3=(-0.26,-1.48)$ are the kernel centers generated with UDR numbers. 
	The parameters of the linear predictor have been computed according to~\eqref{Eqn:appro_K}  and~\eqref{Eqn:appro_C}.\\
	The penalty matrices $\tilde Q$ and $R$ have been selected as $\tilde Q=I_5$, $R=0.1$.   The terminal penalty matrix, the sets $\bar {\mathcal{W}}$, $\mathcal{V}$, the gain matrix $K$, the robust invariant set, and the terminal constraint ${\mathcal{S}}_f$ have been computed similar to that in Section~\ref{sec:example-van}. 
	The prediction horizon has been chosen as $N=10$. \\
	The proposed r-KMPC has  been implemented with $x_0=(0.2,1)$. 
	The simulation results  are reported in Figures~\ref{fig:sta-con-sin-ip}-\ref{fig:sta_con_no_external_dis-ip}. It can be seen from Figures~\ref{fig:sta-con-sin-ip} and~\ref{fig:real-tra-sin-ip}  that 
	the robustness of the closed-loop control system, the satisfaction of state and control constraints as well as the convergence of the nominal system are verified. 
	Also,  under no external perturbation, the state and control of r-KMPC converge asymptotically to the origin, see Figure~\ref{fig:sta_con_no_external_dis-ip}. \\
	\begin{table}[h!tb]
		\centering \caption{Cumulative cost comparison: ``S" and ``SW" stand for ``sinusoidal" and ``Step-wise" respectively; ``InvQuad" stands for ``Inverse quadratic". }
		\label{tab:Tab_com2}
		\vskip 0.2cm
		\scalebox{0.8}{
			\begin{tabular}{ccccccc}
				\hline
				\multicolumn{3}{c}{\multirow{2}{*}{Algorithm}}& R-KMPC &\multicolumn{3}{c}{KMPC~\cite{korda2018linear}}\\
				\cline{5-7}
				\multicolumn{3}{c}{}	& $n_{\psi}=5$ 	& $n_{\psi}=5$  & $n_{\psi}=15$ & $n_{\psi}=25$   \\
				\hline
				\multirow{8}{*}{Cost $J$} 
				&\multirow{3}{*}{Gaussian kernel}&Nominal  & $175$ & $434$ &  $228$&203\\
				&&	S-noise  & $333$ & $695$ &  $407$&367\\
				& &	UDR-noise  & $191$ & $570$ &  $235$&189\\
				& &	SW-noise  & $201$ & $763$ &  $255$&214\\
				\cline{2-7}
				&\multirow{3}{*}{InvQuad kernel}&	Nominal  & $173$ & $512$ &  $252$&--\\
				&&	S-noise  & $319$ & $871$ &  $432$&--\\
				& &	UDR-noise  & $236$ & $546$ &  $276$&--\\
				& &	SW-noise  & $248$ & $484$ &  $261$&--\\
				\hline
			\end{tabular}
		}	 
	\end{table}
	Numerical comparison with KMPC~\cite{korda2018linear}: 
	The comparison has been performed similar to that in Section~\ref{sec:example-van}, from the simulation results in Figure~\ref{fig:sta_con_no_external_dis-ip}, one can observe the diverging trends of the state and control in KMPC and asymptotic convergence of the state and control in our approach. 
	The numerical comparison in terms of the cumulative costs of both approaches in Table~\ref{tab:Tab_com2} show that the proposed r-KMPC results in a lower cost consumption compared with KMPC~\cite{korda2018linear} even using a greater number of observable functions.}
	\subsection{Regulation of a non-affine system}
	Consider a non-affine system~\cite{ge1999adaptive}, whose continuous-time model is
		\begin{equation}\label{Eqn:ct-nonaffine}
	\begin{array}{ll}
	\begin{bmatrix}
	\dot x_1\\
	\dot x_2
	\end{bmatrix}=&\begin{bmatrix}
	x_2\\
	x_1^2+0.15u^3+0.1(1+x_2^2)u+sin(0.1u)
	\end{bmatrix}+w_o,
	\end{array}
	\end{equation}
	where $x=(x_1,x_2)$ are the state variables, $u$ is the control, $w_o={\rm sin} (10\pi t)$. 
	The constraints to be enforced are
	$-2.5\leq x\leq2.5,\ 
	-25\leq u\leq 25.$\\
	\begin{figure}[h]
		\centering
		\includegraphics[width=0.5\textwidth]{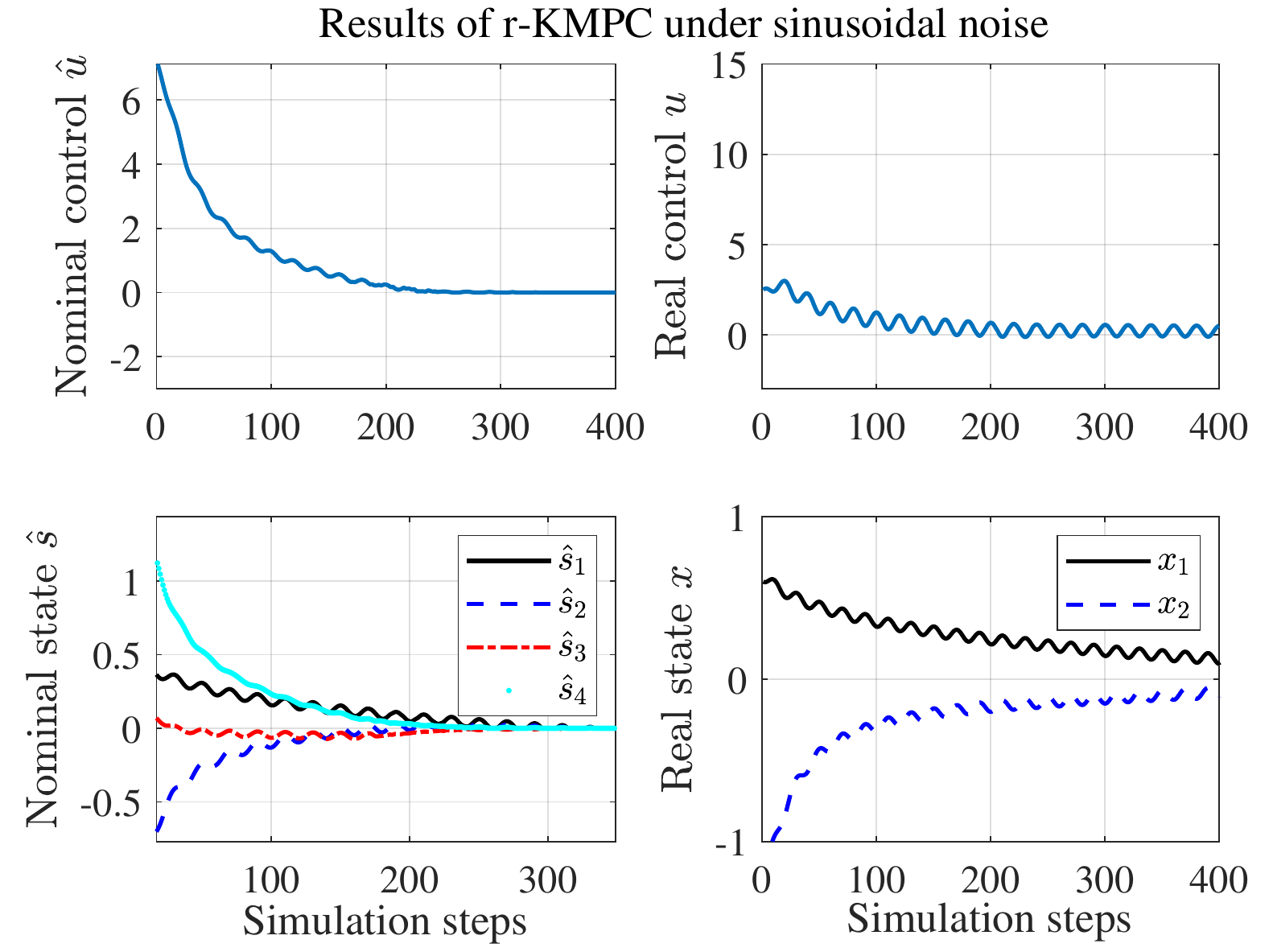} 
		\caption{The  state and control of the controlled nonaffine system with r-KMPC under sinusoidal noise.}
		\label{fig:sta-con-sin-nonaffine}
	\end{figure}
	\begin{figure}[h]
		\centering
		\includegraphics[width=0.5\textwidth]{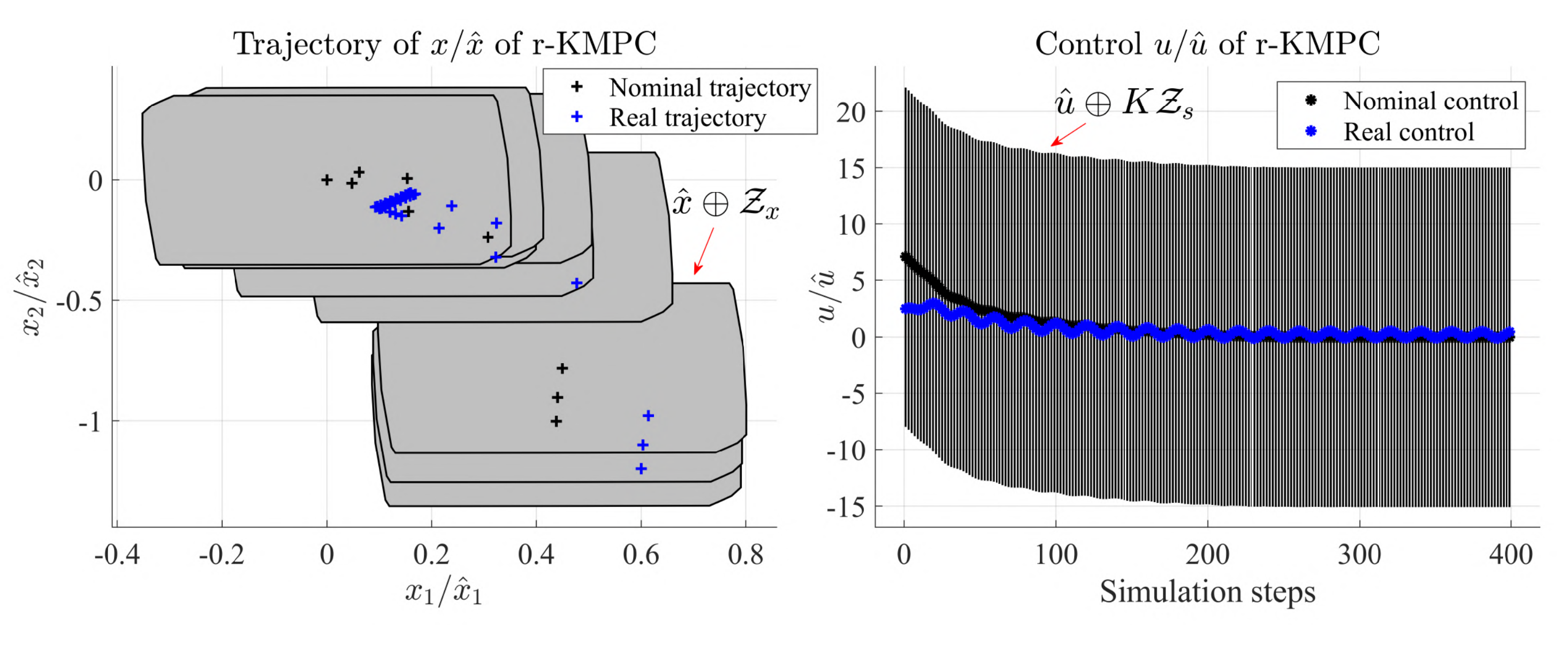}
		\caption{The state and control tubes of the controlled non-affine system under sinusoidal noise. In the left (right) panel, the real trajectory, marked with blue + ($\ast$), lies in the grey tube, i.e., $\mathcal{Z}_x$ ($K\mathcal{Z}_s$), centered at the nominal one, marked with black + ($\ast$).}
		\label{fig:real-tra-sin-nonaffine}
	\end{figure}
	\begin{figure}[h]
		\centering
		\includegraphics[width=0.5\textwidth]{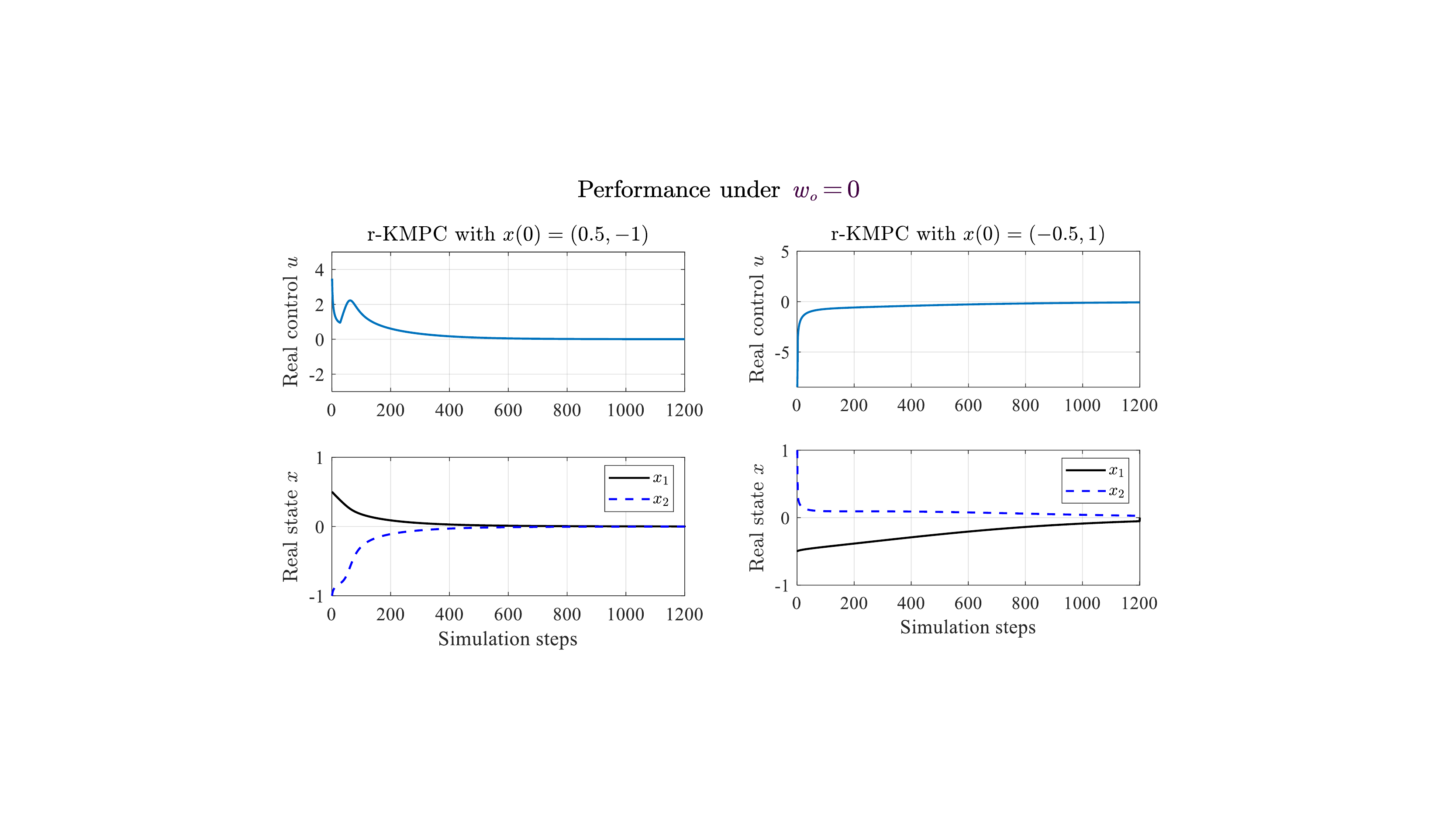} 
		\caption{Performance of the controlled nona-ffine system with r-KMPC under $w_o=0$. The state and control converge asymptotically to the origin.}
		\label{fig:sta_con_no_external_dis-nonaffine}
	\end{figure}
	{\hspace{-1.45mm}To obtain the Koopman model for control, model~\eqref{Eqn:ct-nonaffine} is discretized with $T=0.005s$. The data samples have been obtained with $M=5\cdot 10^4$ under a UDR control policy. The observable functions have been constructed as $\Psi(x)=(x,\psi_1(x),\psi_2(x),\psi_3(x))-(\bm 0,\psi_{1}(0),\psi_{2}(0),\psi_3(0))$, where  $\psi_i(x)=\|x-c_i\|{\rm log}(\|x-c_i\|)$, $i=1,2,3$ are polyharmonic kernels, $n_{\psi}=5$, $c_i$, $i=1,2,3$ are the kernel centers generated with UDR numbers.} 
	The parameters of the Koopman model have been computed according to~\eqref{Eqn:appro_K}   and~\eqref{Eqn:appro_C}.\\
	To control model~\eqref{Eqn:ct-nonaffine} with  the r-KMPC,  the penalty matrices $\tilde Q$, $R$ have been selected as $\tilde Q=I_5$, $R=0.1$, and the prediction horizon has been chosen as $N=30$.  The terminal penalty matrix, the sets $\bar {\mathcal{W}}$, $\mathcal{V}$, the gain matrix $K$, the robust invariant set, and the terminal constraint ${\mathcal{S}}_f$ have been computed similar to that in Section~\ref{sec:example-van}. 
	The proposed r-KMPC has  been implemented with $x_0=(0.6,-1.2)$. 
	The simulation results in Figures~\ref{fig:sta-con-sin-nonaffine} and~\ref{fig:real-tra-sin-nonaffine}  illustrate that 
	the robustness of the closed-loop control system, the satisfaction of state and control constraints as well as the convergence of the nominal system are verified. 	Also,  under no external perturbation, the state and control of r-KMPC converge asymptotically to the origin, see Figure~\ref{fig:sta_con_no_external_dis-nonaffine}. 
		\section{Conclusions}
		{\color{black}
			In this paper,  we proposed a robust tube-based model predictive control scheme, i.e., r-KMPC, for nonlinear systems with additive disturbances as well as state and control constraints.
			The proposed r-KMPC relies upon a lifted global linear model built using Koopman operators. The closed-loop robust controller is composed of a nonlinear control action computed by an online linear MPC using the nominal Koopman model and a nonlinear static state-feedback policy. The robustness of the closed-loop system was verified under internal modeling errors and exogenous disturbances. Moreover, the asymptotic stability of the controlled system under no exogenous disturbance was proven under mild assumptions.  Simulation results verified the effectiveness of the proposed approach.}
		%
		%
		\appendix
		\section{Statistical validations  of sets $\bar{\mathcal{W}}$ and $\mathcal{V}$}\label{sec:app-b}
		{\color{black}	The derivation of $\bar{\mathcal{W}}$ and $\mathcal{V}$ is not straightforward since they show dependencies on $\Psi(x)$ and the datasets used in~\eqref{Eqn:appro_K} (or~\eqref{Eqn:appro_K-modify}) and~\eqref{Eqn:appro_C}. In the following, we show that the above sets can be estimated using statistic learning theory. As described in \cite{hertneck2018learning}, the true risk of a learning problem can be estimated according to the law of large numbers. This relies upon a relaxed risk evaluation condition developed by Hoeffding $(1963)$:
			\begin{equation}\label{Eqn:risk-eva}
			P\left(\left|\hat G(f_l)-G(f_l)\right| \geq \epsilon\right) \leq \delta_r\end{equation}
			where $\delta_r:=2 \exp \left(-2 L \epsilon^{2}\right)$, $\hat G(f_l)$ and $G(f_l)$ are the empirical and true risk respectively, $f_l$ is the learned function, and $L$ is the number of  samples. For fixed $f_l$, provided a large number of datasets, the empirical risk is a good estimate of the true one.\\
	\xl{
			{\emph {Loss function for validation:}
				Let $L$ tuples $(u_i,\hat w_{o,i},x_i,x_i^+)$ be generated under Assumption~\ref{Eqn:data-distribution}. Let $\bar w_i=\Psi(x_i^{+})-\hat s_i^{+}$ where $\hat s_i^{+}$ is computed by~\eqref{Eqn:unpert} with $\hat s=\Psi(x_i)$ and $\hat u=u_i$ and let $v_i=x_i-C\Psi(x_i)$. We define the loss function as 
				\begin{equation}\label{Eqn:loss-set-va} 
				\ell_{\star}(u_i,x_i,x_i^+)=\left\{\begin{array}{ll}{0} & {\text { if  $g_{\star}=1$}} \\ {1} & {\text { otherwise }}\end{array}\right.
				\end{equation}
				where $\star=\bar w,v$ in turns, $g_{\bar w_i}=1$ stands for $\bar w_i\in\bar{\mathcal{W}}$ and $g_{v_i}=1$ stands for $v_i\in\mathcal{V}$.\\ 
					With~\eqref{Eqn:loss-set-va}, for any $\star=\bar w, v$, the empirical risk of the learned linear predictor is $\hat G_{\star}=\frac{1}{L}\sum_{i=1}^L\ell_{\star}(u_i,x_i,x_i^+)$. In view of~\eqref{Eqn:risk-eva}, $P_{\star}(\ell=0)=G_{\star}\leq \hat G_{\star}+\epsilon_{\star}$. For the validation of uncertainty sets, we choose  $\bar G_{\star}\geq G_{\star}$ and $\delta_r>0$, then we check if for the $L$ samples that~\cite{hertneck2018learning} 
	\begin{equation}\label{Eqn:validation}
\bar G_{\star}\geq\hat G_{\star}+\epsilon_{\star}
\end{equation} with a confidence level of $1-\delta_r$, where $\epsilon_{\star}=\sqrt{-{\log \left({0.5\delta_{r}}\right)}/{2 L}}$.\\
					\begin{algorithm}[h]
						\caption{Computation of $\bar {\mathcal{W}}$ and $\mathcal{V}$}
						\label{alg:com-sets}		
					\begin{algorithmic}
						\State \textbf{1)} Set initial sets  $\bar {\mathcal{W}}_I$ and $\mathcal{V}_I$;
						\State \textbf{2)} Set $\bar G_{\star}\geq G_{\star}$ and $\delta_r$, we calculate  $\hat G_{\star}$ with~\eqref{Eqn:loss-set-va} and $\epsilon_{\star}=\sqrt{-{\log \left({0.5\delta_{r}}\right)}/{2 L}}$, for $\star=\bar w,v$;
							\If{Condition~\eqref{Eqn:validation} is satisfied for $\star=\bar w$ (or $\star=v$)}
							\State \textbf{3)} Return $\bar {\mathcal{W}}_I$ (or $ \mathcal{V}_I$);
							\Else
							\State \textbf{4)} Increase $L$ or enlarge the size of ${\bar {\mathcal{W}}}_I$ (or ${\bar {\mathcal{V}}}_I$);
							\State \textbf{5)} Go back to step 2);
							\EndIf
							\State \textbf{6)} Set $\bar{\mathcal{W}}=\gamma_w{\bar {\mathcal{W}}}_I$ and $\mathcal{V}=\gamma_v {\mathcal{V}}_I$ where  $\gamma_w,\gamma_v>1$.
						\end{algorithmic}
					\end{algorithm}}
				The computation steps of $\bar {\mathcal{W}}$ and $\mathcal{V}$ are given in Algorithm~\ref{alg:com-sets}. The rationale behind the uncertainty scaling in step 6) of Algorithm 2 is given in the following proposition.
Denote  $\mathcal{D}_{h}=\{h_i\}_{i=1}^L$, $h_i=x_i,u_i,\hat w_{o,i}$ in turns; define $d_{x}=\max_{x,y\in\mathcal{D}_x,y\in x_{\mathcal{N}}}\|x-y\|$, $d_{u}=\max_{u\in\mathcal{D}_u,y\in y_{\mathcal{N}}}\|u-y\|$, $d_{\hat w}=\max_{w\in\mathcal{D}_{w_o},y\in u_{\mathcal{N}}}\|w-y\|$, where $z_{\mathcal{N}}$ is the collection of neighbor point of $z$ for $z=x,u,w$. Let $d_{\delta w}$ be the maximal estimation deviation of the real disturbance and the estimated one.}   The following proposition can be stated.
			\begin{proposition}[Uncertainty set scaling]\label{prop:set-cound}
			The real uncertainty sets $\bar{\mathcal{W}}$, $\mathcal{V}$ satisfy $\bar{\mathcal{W}}\subseteq {\bar{\mathcal{W}}}_I\oplus\Delta\mathcal{W}$, ${\mathcal{V}}\subseteq {{\mathcal{V}}}_I\oplus\Delta\mathcal{V}$, where $\Delta\mathcal{W}=\{w\in\mathbb{R}^{n_{\psi}}|\|w\|\leq L_s L_{\Psi}d_x+L_ud_{u}+L_{\delta w} d_{\delta w}+
			L_{\hat w}d_{\hat w}\}$, $\Delta\mathcal{V}=\{v\in\mathbb{R}^{ n}|\|v\|\leq L_v d_x\}$, $L_{\Psi}$ and $L_v$ are the local Lipschitz constants of $\Psi(x)$ and $x-\Psi^{-1}$ respectively.
					\end{proposition}
				\textbf{Proof}. Let $w_1\in{\bar{\mathcal{W}}}_I\ominus D\hat{\mathcal{W}}_o$, $w_2\in\bar{\mathcal{W}}\backslash ({\bar{\mathcal{W}}}_I\ominus D\hat{\mathcal{W}}_o)$. In view of~\eqref{Eqn:lipsch-1}, one has that $
				\|w_1-w_2\|\leq L_s L_{\Psi} d_x+
				L_u d_u+ L_{\delta w} d_{\delta w}+
				L_{\hat w}d_{\hat w}
				$
				where the second inequality follows from the Lipschitz continuity of $\Psi(x)$. Hence, $\bar{\mathcal{W}}\subseteq {\bar{\mathcal{W}}}_I\oplus\Delta\mathcal{W}$.
				Likewise, let $v_1\in{{\mathcal{V}}}_I$, $v_2\in\mathcal{V}\backslash {{\mathcal{V}}}_I$, then one has that 
			    $
				\|v_1-v_2\|\leq  L_v d_x.
				$ Hence, ${\mathcal{V}}\subseteq {{\mathcal{V}}}_I\oplus\Delta\mathcal{V}$.
			\hfill $\square$
}

					\bibliographystyle{unsrt}
					\bibliography{refletter}
			
			\end{document}